\documentclass[english]{article}
\usepackage[]{fontenc}
\usepackage[latin1]{inputenc}
\usepackage{amsmath}
\usepackage{amssymb}

\makeatletter

\newcommand{\noun}[1]{\textsc{#1}}

\usepackage{geometry}

\usepackage{graphicx}

\usepackage{amsfonts}



\usepackage{babel}
\makeatother
\begin{document}

\title{\textbf{Quantization of (}$\mathbf{2+1}$\textbf{)-spinning particles
and bifermionic constraint problem} }

\author{\noun{R. Fresneda}\thanks  {E-mail: fresneda@fma.if.usp.br },
S.P. Gavrilov\thanks  {Dept. F\'{\i}sica e Qu\'{\i}mica, UNESP,
Campus de Guaratinguet\'{a}, Brazil; on leave from Tomsk State Pedagogical
University, 634041 Tomsk, Russia; email: gavrilovsp@hotmail.com},
D.M. Gitman\thanks  {E-mail: gitman@dfn.if.usp.br}, and P.Yu. Moshin\thanks  {On
leave from Tomsk State Pedagogical University, 634041 Tomsk, Russia;
e-mail: moshin@dfn.if.usp.br}\\
 \\
Instituto de F\'{\i}sica, Universidade de S\~{a}o Paulo,\\
Caixa Postal 66318-CEP, 05315-970 S\~{a}o Paulo, S.P., Brazil  }

\maketitle
\begin{abstract}
This work is a natural continuation of our recent study in quantizing
relativistic particles. There it was demonstrated that, by applying
a consistent quantization scheme to the classical model of a spinless
relativistic particle as well as to the Berezin-Marinov model of $3+1$
Dirac particle, it is possible to obtain a consistent relativistic
quantum mechanics of such particles. In the present article we apply
a similar approach to the problem of quantizing the massive $2+1$
Dirac particle. However, we stress that such a problem differs in
a nontrivial way from the one in $3+1$ dimensions. The point is that
in $2+1$ dimensions each spin polarization describes different fermion
species. Technically this fact manifests itself through the presence
of a bifermionic constant and of a bifermionic first-class constraint.
In particular, this constraint does not admit a conjugate gauge condition
at the classical level. The quantization problem in $2+1$ dimensions
is also interesting from the physical viewpoint (e.g. anyons). In
order to quantize the model, we first derive a classical formulation
in an effective phase space, restricted by constraints and gauges.
Then the condition of preservation of the classical symmetries allows
us to realize the operator algebra in an unambiguous way and construct
an appropriate Hilbert space. The physical sector of the constructed
quantum mechanics contains spin-$1/2$ particles and antiparticles
without an infinite number of negative-energy levels, and exactly
reproduces the one-particle sector of the $2+1$ quantum theory of
a spinor field.
\end{abstract}

\section{Introduction}

This work is a natural continuation of our recent study \cite{GavGi00a,GavGi00b,GavGi01}
which was devoted to the consistent quantization of classical and
pseudoclassical models of relativistic particles. We recall that in
the first article \cite{GavGi00a} it is demonstrated that, by applying
the consistent quantization scheme (canonical quantization, combined
with the analysis of constraints and symmetries, as well as of the
physical sector) to the classical model of a spinless relativistic
particle, it is possible to obtain a consistent relativistic quantum
mechanics (QM) of such a particle. Remarkably, the problem of the
infinite number of energy levels and of the indefinite metric is solved
in the same manner as in the corresponding quantum field theory (QFT),
i.e., by properly defining the physical sector of the Hilbert space.
In external backgrounds that do not violate vacuum stability, the
constructed QM turns out to be completely equivalent to the one-particle
sector of the QFT. We stress that the Schrödinger equation of the
QM is equivalent to a pair of relativistic wave equations (in this
particular case to the pair of Klein-Gordon equations), namely, to
an equation for a particle with charge $q$ and to an equation for
an antiparticle with charge $-q$.

As a logical extension of the approach \cite{GavGi00a}, we consider
the quantization of pseudoclassical models of spinning particles.
In this relation one ought to recall that there does not exist a unique
framework for the description of relativistic spinning particles which
embraces all possible cases: particles with integer and half-integer
spin, massive particles and massless particles, particles in even
and in odd dimensions. In all these cases, the action's structure
differs in an essential manner, and therefore each case requires a
completely different treatment in the course of its quantization.
In our works \cite{GavGi00b,GavGi01}, we started the quantization
program of spinning particles considering the pseudoclassical Berezin-Marinov
\cite{BM} action of the massive $3+1$ Dirac particle, the structure
of which is typical for all massive relativistic particles with half-integer
spin in even dimensions. The constraint structure of this model in
the Hamiltonian formulation allows one to fix completely the gauge
freedom at the classical level. In spite of the essential technical
difficulties involving the realization of the commutation relations
and the Hamiltonian construction, one can carry out the canonical
quantization scheme leading to the consistent (as in the spinless
case) QM of the $3+1$ Dirac particle.

In the present article we consider the problem of quantizing the massive
$2+1$ Dirac particle using the pseudoclassical model first proposed
by Gitman, Gonçalves, and Tyutin (GGT) in \cite{GitGoTy}. From this
particular model, one can devise the general quantization scheme for
half-integer spinning particles in odd dimensions \cite{GitTy97},
since the action for the general case has the same essential structure
as that of the GGT model. More remarks are in order regarding the
choice of the model for the massive $2+1$ Dirac particle. Namely,
we note that a number of alternative models have been proposed for
the description of a spinning particle in $2+1$ dimensions (see,
e.g., \cite{GitGoTy,P}). We also note that in $2+1$ dimensions,
a direct dimensional reduction of the Berezin-Marinov action does
not reproduce the minimal quantum theory of spinning particles, which
must provide only one spin projection value ($1/2$ or $-1/2$). In
the papers \cite{P,CPV93}, two modifications were proposed. One of
them is not minimal and is $P$- and $T$- invariant, so that an anomaly
is present. The other does not possess the desirable gauge supersymmetries.
The GGT action is gauge supersymmetric and reparametrization invariant.
Furthermore, it is $P$- and $T$-noninvariant, in accordance with
the expected properties of the minimal theory in $2+1$ dimensions.

We stress that the quantization of the GGT model differs in a nontrivial
way from the one presented for the Berezin-Marinov model. The point
is that in $2+1$ dimensions (as well as in any odd-dimensional case)
to each spin projection corresponds a different particle, because
distinct spin projections of a $2+1$ spinning particle belong to
distinct irreducible representations, and thus describe different
fermion species. Technically this fact manifests itself in the model
through the presence of a bifermionic constant and of a bifermionic
first-class constraint. This constraint does not admit a conjugate
gauge condition at the classical level. However, since the corresponding
operator has a compact spectrum, it can be consistently used to fix
the remaining gauge freedom at the quantum level according to Dirac.
Such problems do not appear in the Berezin-Marinov case. Our interest
in the GGT model does not reside entirely upon these mentioned departures
from the Berezin--Marinov model. The quantization problem of a particle
in $2+1$ dimensions is a very interesting one from the physical viewpoint.
There is a direct relation to field theory in $2+1$ dimensions \cite{b1,b2},
which has recently attracted much attention, due to non-trivial topological
properties, and especially due to the possible existence of particles
with fractional spin and exotic statistics (anyons). There is also
a strong relation of the $2+1$ quantum theory to the fractional Hall
effect, high-$T_{c}$ superconductivity, etc. \cite{b1}. Thus, we
hope to have motivated the construction of a consistent relativistic
QM of a spinning particle in $2+1$ dimensions.

The paper is organized as follows. In Section 2, we study the classical
properties of the given pseudoclassical model and present its detailed
Hamiltonian analysis. We focus on the selection of the physical degrees
of freedom and on an adequate gauge-fixing. We obtain a Hamiltonian
formulation of the model in an effective phase space, restricted by
constraints and gauges. We gauge-fix two of the initial gauge symmetries,
and retain an effective bifermionic first-class constraint, which
does not admit gauge-fixing. In Section 3, we apply a quantization
approach, being a combination of the canonical and the Dirac schemes,
in which the bifermionic first-class constraint is imposed at the
quantum level to select admissible state-vectors. We present a detailed
construction of the Hilbert space. Then we reformulate the time-evolution
in terms of the physical time, and verify that the constructed theory
has the necessary symmetry properties. We select a physical sector
which describes the consistent relativistic QM of particles in $2+1$
dimensions without an infinite number of negative-energy levels. In
Section 4, we make a comparison of the constructed QM with the one-particle
sector of the $2+1$ QFT. In Section 5, we summarize all the results
obtained in our paper. In the Appendix, we justify the selected Hamiltonian
realization, considering the semiclassical limit of the QM constructed.

\section{Pseudoclassical model and its constraint structure}

\subsection{Lagrangian and Hamiltonian formulations}

In order to describe classically (that is, pseudoclassically) massive
relativistic spin-$1/2$ charged particles in $2+1$ dimensions, we
take the action first proposed in \cite{GitGoTy}. It has the form
\begin{align}
 & S=\int_{0}^{1}L\, d\tau\,,\; L=-\frac{z^{2}}{2e}-e\frac{m^{2}}{2}-q\dot{x}^{\mu}A_{\mu}+ieqF_{\mu\nu}\xi^{\mu}\xi^{\nu}-im\xi^{3}\chi-\frac{1}{2}\theta m\kappa-i\xi_{n}\dot{\xi}^{n}\,,\nonumber \\
 & z^{\mu}=\dot{x}^{\mu}-i\xi^{\mu}\chi+i\varepsilon^{\mu\nu\lambda}\xi_{\nu}\xi_{\lambda}\kappa\,.\label{2.1}\end{align}
 Here $e$, $\kappa,$ and $x^{\mu}$, $\mu=0,1,2$, are even variables,
while $\chi$ and $\xi^{n}$, $n=(\mu,3)$, are odd variables; the
Minkowski metric in $2+1$ dimensions reads $\eta_{\mu\nu}=\mathrm{diag}(1,-1,-1)$,
and in $3+1$ dimensions is $\eta_{mn}=\mathrm{diag}(1,-1,-1,-1)$;
$\theta$ is an even constant; $\varepsilon^{\lambda\mu v}$ is the
Levi-Civita tensor in $2+1$ dimensions normalized as $\varepsilon^{012}=1$,
and summation over repeated indices is assumed. The particle interacts
with an arbitrary external gauge field $A_{\mu}(x)$, which can be
of Maxwell and/or Chern-Simons nature, $F_{\mu\nu}$ $=\partial_{\mu}A_{\nu}-\partial_{\nu}A_{\mu}$
is the strength tensor of this field, and $q$ is the $U(1)$-charge
of the spinning particle. We assume that the coordinates $x^{\mu}$
and $\xi^{\mu}$ are $2+1$ Lorentz vectors; $e$, $\kappa$, $\xi^{3}$,
and $\chi$ are Lorentz scalars. All the variables depend on the parameter
$\tau\in[0,1]$, which plays here the role of time. Dots above the
variables denote their derivatives with respect to $\tau$. The action
(\ref{2.1}) is invariant under the restricted Lorentz transformations,
but is $P$- and $T$-noninvariant, in accordance with the expected
properties of the minimal theory in $2+1$ dimensions.

We recall that the action is invariant under the reparametrizations
\[
\delta x^{\mu}=\dot{x}^{\mu}\varepsilon\,,\;\;\delta e=\frac{d}{d\tau}\left(e\varepsilon\right)\,,\;\;\delta\xi^{n}=\dot{\xi}^{n}\varepsilon\,,\;\;\delta\chi=\frac{d}{d\tau}\left(\chi\varepsilon\right)\,,\;\;\delta\kappa=\frac{d}{d\tau}\left(\kappa\varepsilon\right)\,,\]
 where $\varepsilon\left(\tau\right)$ is an even gauge parameter,
and under two types of gauge supertransformations \begin{align*}
 & \delta x^{\mu}=i\xi^{\mu}\epsilon\,,\;\;\delta e=i\chi\epsilon\,,\;\;\delta\xi^{\mu}=\frac{z^{\mu}}{2e}\epsilon\,,\;\;\delta\xi^{3}=\frac{m}{2}\epsilon\,,\;\;\delta\chi=\dot{\epsilon\,},\;\;\delta\kappa=0\,,\\
 & \delta x^{\mu}=-i\varepsilon^{\mu\nu\lambda}\xi_{\nu}\xi_{\lambda}\vartheta\,,\;\;\delta\xi^{\mu}=\frac{1}{e}\varepsilon^{\mu\nu\lambda}z_{\nu}\xi_{\lambda}\vartheta\,,\;\;\delta\kappa=\dot{\vartheta}\,,\;\;\delta e=\delta\xi^{3}=\delta\chi=0\,,\end{align*}
 where $\epsilon(\tau)$ is an odd gauge parameter and $\vartheta(\tau)$
is an even gauge parameter.

We note that $e$, $\chi$, and $\kappa$ are degenerate coordinates,
since their time derivatives are not present in the action. In what
follows, we consider a reduced hamiltonization scheme for theories
with degenerate coordinates \cite{GitTy}, in which momenta conjugate
to the degenerate coordinates are not introduced. To proceed with
the hamiltonization, we introduce the velocities $\upsilon^{\mu}$
and $\alpha^{\mu}$ and write the action (\ref{2.1}) in the first-order
formalism as \[
S^{\upsilon}=\int_{0}^{1}\left[L^{\upsilon}+p_{\mu}\left(\dot{x}^{\mu}-\upsilon^{\mu}\right)+\pi_{n}\left(\dot{\xi}^{n}-\alpha^{n}\right)\right]d\tau=\int_{0}^{1}\left(p_{\mu}\dot{x}^{\mu}+\pi_{n}\dot{\xi}^{n}-H^{\upsilon}\right)d\tau\,,\]
 where \begin{align*}
 & L^{\upsilon}=-\frac{\bar{z}^{2}}{2e}-e\frac{m^{2}}{2}-q\upsilon^{\mu}A_{\mu}+iqeF_{\mu\nu}\xi^{\mu}\xi^{\nu}-im\xi^{3}\chi-\frac{1}{2}\theta m\kappa-i\xi_{n}\alpha^{n}\,,\\
 & \bar{z}^{\mu}=\upsilon^{\mu}-i\xi^{\mu}\chi+i\varepsilon^{\mu\nu\lambda}\xi_{\nu}\xi_{\lambda}\kappa\,,\;\;\; H^{\upsilon}=p_{\mu}\upsilon^{\mu}+\pi_{n}\alpha^{n}-L^{\upsilon}.\end{align*}
 The variables $p_{\mu}$ and $\pi_{n}$ should be treated as conjugate
momenta to the coordinates $x^{\mu}$ and $\xi^{n}$ , respectively.
The ordering of variables in the Hamiltonian $H^{\upsilon}$ complies
with the usual convention for the choice of derivatives with respect
to coordinates as right-hand ones and those with respect to momenta
as left-hand ones.

The equations of motion with respect to the velocities and the degenerate
coordinates read \begin{align*}
 & \frac{\delta S^{\upsilon}}{\delta\upsilon^{\mu}}=-p_{\mu}-\frac{1}{e}\bar{z}_{\mu}-qA_{\mu}=0,\;\;\frac{\delta S^{\upsilon}}{\delta e}=\frac{\bar{z}^{2}}{2e^{2}}-\frac{m^{2}}{2}+iqF_{\mu\nu}\xi^{\mu}\xi^{\nu}=0,\\
 & \frac{\delta S^{\upsilon}}{\delta\kappa}=-\frac{i}{e}\bar{z}^{\mu}\varepsilon_{\mu\nu\lambda}\xi^{\nu}\xi^{\lambda}-\frac{1}{2}\theta m=0,\;\;\frac{\delta S^{\upsilon}}{\delta\alpha^{a}}=\pi_{n}+i\xi_{n}=0,\\
 & \frac{\delta S^{\upsilon}}{\delta\chi}=\frac{1}{e}\left(\upsilon^{\mu}\xi_{\mu}+i\varepsilon^{\mu\nu\lambda}\xi_{\mu}\xi_{\nu}\xi_{\lambda}\kappa\right)-im\xi^{3}=0.\end{align*}
 The equations $\delta S^{\upsilon}/\delta\alpha^{n}=0$ lead to the
primary constraints \begin{equation}
\varphi_{n}=\pi_{n}+i\xi_{n}\,,\label{2.2}\end{equation}
 and the equations $\delta S^{\upsilon}/\delta\upsilon^{\mu}=0$ can
be used to express the velocities $\upsilon^{\mu}$, viz., \begin{equation}
\upsilon^{\mu}=-e\left(p^{\mu}+qA^{\mu}\right)+i\xi^{\mu}\chi-i\varepsilon^{\mu\nu\lambda}\xi_{\nu}\xi_{\lambda}\kappa\,.\label{2.3}\end{equation}
 Substituting this relation into the other equations, we obtain more
primary constraints \begin{align}
\phi_{1} & =\left(p+qA\right)^{\mu}\xi_{\mu}+m\xi^{3},\nonumber \\
\phi_{2} & =\left(p+qA\right)^{2}-m^{2}+2iqF_{\mu\nu}\xi^{\mu}\xi^{\nu},\nonumber \\
\phi_{3} & =\varepsilon_{\mu\nu\lambda}\left(p+qA\right)^{\mu}\xi^{\nu}\xi^{\lambda}+\frac{i}{2}\theta m\,.\label{2.4}\end{align}

Upon substituting (\ref{2.3}) into the Hamiltonian $H^{\upsilon}$,
we get the total Hamiltonian \begin{equation}
H^{\left(1\right)}=\Lambda_{1}\phi_{1}+\Lambda_{2}\phi_{2}+\Lambda_{3}\phi_{3}+\lambda^{n}\varphi_{n}\,,\label{2.5}\end{equation}
 where $\Lambda_{1}=-i\chi\,$, $\Lambda_{2}=-e/2$, $\,\,\Lambda_{3}=-i\kappa$,
and $\,\lambda^{n}=-\alpha^{n}$ are henceforth Lagrange multipliers.
One can see that the total Hamiltonian is proportional to the constraints.
The resulting dynamically equivalent action is \[
S^{\left(1\right)}=\int_{0}^{1}\left(p_{\mu}\dot{x}^{\mu}+\pi_{n}\dot{\xi}^{n}-H^{\left(1\right)}\right)d\tau\,.\]

In the following, we shall make use of the Dirac brackets with respect
to a set $\varphi$ of second-class constraints, \[
\left\{ A,B\right\} _{D\left(\varphi\right)}=\left\{ A,B\right\} -\left\{ A,\varphi_{a}\right\} C^{ab}\left\{ \varphi_{b},B\right\} .\]
 Here $C^{ab}\left\{ \varphi_{b},\varphi_{c}\right\} =\delta_{c}^{a}$
and the Poisson brackets of the functions $F$ and $G$ of definite
Grassmann parities $\varepsilon(F)$ and $\varepsilon\left(G\right)$
are given by \begin{equation}
\left\{ F,G\right\} =\frac{\partial F}{\partial x^{\mu}}\frac{\partial G}{\partial p_{\mu}}+\frac{\partial F}{\partial\xi^{n}}\frac{\partial G}{\partial\pi_{n}}-\left(-1\right)^{\varepsilon\left(F\right)\varepsilon\left(G\right)}(F\leftrightarrow G).\label{2.6}\end{equation}

One ought to say that in the classical theory the quantity $\theta$
is a bifermionic constant. In the quantum theory, though, it turns
out to be a real number (see Sect. 5 Discussion).

\subsection{Constraint reorganization and gauge-fixing}

In analogy to \cite{GavGi00b,GavGi01}, we first reorganize the constraints
(\ref{2.2}) and (\ref{2.4}) into an equivalent set $(\mathbf{T},\varphi)$
such that $\mathbf{T=}\left(T_{1},T_{2},T^{\prime}\right)$ is a set
of first-class constraints, and $\varphi$ is a set of second-class
constraints, \[
\left.\left\{ \mathbf{T},\mathbf{T}\right\} \right|_{\mathbf{T},\varphi=0}=\left.\left\{ \mathbf{T},\varphi\right\} \right|_{\mathbf{T},\varphi=0}=0\,.\]
 The new constraints $\mathbf{T}$ have the form \begin{align}
T_{1} & =\left(p+qA\right)_{\mu}\left(\pi^{\mu}-i\xi^{\mu}\right)+m\left(\pi^{3}-i\xi^{3}\right)\,,\nonumber \\
T_{2} & =p_{0}+qA_{0}+\zeta r\,,\;\;\zeta=-\mathrm{sgn}\left[p_{0}+qA_{0}\right]\,,\nonumber \\
T^{\prime} & =\varepsilon_{\mu\nu\lambda}\left(p+qA\right)^{\mu}\xi^{\nu}\pi^{\lambda}+\frac{1}{2}\theta m\,,\label{2.7}\end{align}
 where $\zeta=\pm1$, and $r=\sqrt{m^{2}+\left(p_{k}+qA_{k}\right)^{2}+2qF_{\mu\nu}\xi^{\mu}\pi^{\nu}}$
is the principal value of the square root%
\footnote{We define the principal value of the square root of an expression
containing Grassmann variables as the one which is positive when the
generating elements of the Grassmann algebra are set to zero.%
}. One can see that $T_{2}$ and $\phi_{2}$ are related as \[
\phi_{2}=-2\zeta rT_{2}-\frac{i}{2}\varphi_{n}\eta^{n\tilde{n}}\left\{ \varphi_{\tilde{n}},\phi_{2}\right\} +\left(T_{2}\right)^{2}\,.\]

In terms of the $\mathbf{T}$-constraints, the Hamiltonian (\ref{2.5})
becomes \begin{equation}
H^{\left(1\right)}=\Lambda_{1}T_{1}+\Lambda_{2}T_{2}+\Lambda^{\prime}T^{\prime}\,,\label{2.8}\end{equation}
 with redefined Lagrange multipliers.

Our goal is to quantize this theory, so supplementary gauge conditions
will be imposed upon the first-class constraints $T_{1}$ and $T_{2}$,
except the constraint $T^{\prime}$, which is of bifermionic nature.
The problem related to its gauge fixing is still open (see discussion
in \cite{GitTy97,GGT94,bifermionic,GalG99}). In the end, there will
remain a first-class constraint $T^{\prime}$ reduced on the constraint
surface, while the second-class set of all the other constraints and
gauge conditions will be used to construct Dirac brackets. The surviving
first-class constraint will be enforced on state vectors of the quantized
theory in order to fix the remaining gauge freedom according to Dirac.

We impose the following gauge-fixing conditions \begin{align}
\phi_{1}^{G} & =\pi^{0}-i\xi^{0}+\zeta\left(\pi^{3}-i\xi^{3}\right),\label{2.9}\\
\phi_{2}^{G} & =x_{0}-\zeta\tau\,,\;\zeta=\pm1.\label{2.10}\end{align}
 The gauge (\ref{2.9}), chosen to fix the gauge freedom related to
$T_{1}$, reduces the set of independent spin variables. The gauge
(\ref{2.10}), chosen to fix the gauge freedom related to $T_{2}$,
is the chronological gauge \cite{GitTy90a,GavGi00a,GavGi00b}. The
resulting set of constraints $(T_{1},T_{2},\phi_{1}^{G},\phi_{2}^{G},\varphi)$
is second-class.

In order to simplify the Poisson brackets between constraints, and
to write the dynamics in terms of independent variables, we pass to
a set $\left(\Phi,\tilde{T}\right)$ which is equivalent to the set
of constraints $(\mathbf{T},\phi_{1}^{G},\phi_{2}^{G},\varphi).$
Here $\Phi$ are the second-class constraints \begin{align}
\Phi_{1} & =p_{0}+qA_{0}+\zeta\tilde{\omega},\;\Phi_{2}=\phi_{2}^{G}\,,\;\Phi_{3}=\varphi_{1}\,,\;\Phi_{4}=\varphi_{2}\,,\nonumber \\
\Phi_{5} & =-\frac{i}{2}T_{1}+bT_{2}+c\phi_{2}^{G}\,,\;\;\;\Phi_{6}=\phi_{1}^{G}\,,\;\;\;\Phi_{7}=\varphi_{0},\;\;\;\Phi_{8}=\varphi_{3}\,,\label{2.11}\end{align}
 where \begin{align*}
 & \tilde{\omega}=\sqrt{\tilde{\omega}_{0}^{2}+\frac{2\zeta qF_{k0}}{\tilde{\omega}_{0}+m}\left(p+qA\right)_{l}\left(\xi^{k}\pi^{l}+\pi^{k}\xi^{l}\right)}\,,\\
 & \tilde{\omega}_{0}=\sqrt{m^{2}+\left(p_{k}+qA_{k}\right)^{2}+2qF_{ik}\xi^{i}\pi^{k}}\,,\\
 & b=\frac{i}{2}\frac{\left\{ \phi_{2}^{G},T_{1}\right\} }{\left\{ \phi_{2}^{G},T_{2}\right\} },\;\;\; c=-\frac{\left\{ -\frac{i}{2}T_{1}+bT_{2},\Phi_{1}\right\} }{\left\{ \phi_{2}^{G},\Phi_{1}\right\} },\end{align*}
 and $\tilde{T}$ reads \[
\tilde{T}=T-\frac{\left\{ T,\Phi_{5}\right\} }{\left\{ \Phi_{6},\Phi_{5}\right\} }\Phi_{6}\,,\;\; T=\left.T^{\prime}\right|_{\Phi=0}\,.\]
 The constraint $\tilde{T}$ is first-class, so it is orthogonal to
all the second-class constraints, $\left.\left\{ \tilde{T},\Phi\right\} \right|_{\Phi=0}=0\,.$

However, for the further consideration, it is convenient to use the
constraint $T,$ which coincides with $\tilde{T}$ and $T^{\prime}$
on the constraint surface, \begin{equation}
T=-m\zeta\left(\xi^{1}\pi^{2}-\xi^{2}\pi^{1}\right)+\frac{1}{2}\theta m\,.\label{2.12}\end{equation}
 The constraint $T,$ which is responsible for parity violation with
respect to reflection of one of the coordinate axes, does not depend
on the space-time coordinates and momenta. The corresponding operator
can be realized as a finite matrix. Therefore its spectrum is compact,
and we thus do not expect standard difficulties with the Dirac quantization
in such a case.

The nonzero Poisson brackets (taken on the constraint surface $\Phi=0$)
between second-class constraints are \begin{align*}
 & \left\{ \Phi_{2},\Phi_{1}\right\} =-\left\{ \Phi_{1},\Phi_{2}\right\} =1,\;\;\left\{ \Phi_{3},\Phi_{3}\right\} =\left\{ \Phi_{4},\Phi_{4}\right\} =-2i\,,\\
 & \left\{ \Phi_{5},\Phi_{6}\right\} =\left\{ \Phi_{6},\Phi_{5}\right\} =\zeta\left(\tilde{\omega}_{0}+m\right),\;\;\left\{ \Phi_{7},\Phi_{7}\right\} =-\left\{ \Phi_{8},\Phi_{8}\right\} =2i\,.\end{align*}

In terms of the new constraints, the Hamiltonian (\ref{2.8}) becomes
\[
H^{\left(1\right)}=\Lambda_{a}\Phi_{a}+\Lambda\tilde{T}\,,\;\;\; a=1,2,5,6.\]
 with redefined Lagrange multipliers.

\subsection{Effective dynamics in the reduced space}

Evidently, not all variables are independent, due to the presence
of constraints. In fact, it is possible to reduce the number of the
variables using some of the second-class constraints. In doing so,
we retain the following set of variables \begin{equation}
\boldsymbol\eta=\left(x^{k},p_{k},\xi^{k},\pi_{k},\zeta\right),\;\;\; k=1,2\,.\label{2.13}\end{equation}
 We hereafter refer to these variables as the basic variables. All
the initial phase-space variables can be expressed in terms of these
basic variables. We remark that the basic variables are not independent,
since there still are constraints among them.

Therefore, we seek to write the dynamics in the reduced space determined
by the evolution of the basic variables (\ref{2.13}) alone. Despite
the fact that the corresponding constraints are time-dependent, it
is still possible to write the evolution equation for the basic variables
by means of Dirac brackets if we introduce a momentum $\epsilon$
canonically conjugate to the time-evolution parameter $\tau$ as was
done in \cite{GitTybk}. This equation reads \begin{equation}
\dot{\boldsymbol\eta}=\left\{ \boldsymbol\eta,\Lambda\tilde{T}+\epsilon\right\} _{D\left(\Phi\right)}\,,\;\;\Phi=0,\;\;\tilde{T}=0,\label{2.14}\end{equation}
 where the Dirac brackets $\{\,,\,\}_{D(\Phi)}$ are constructed with
respect to the constraints (\ref{2.11}).

The new set of constraints $\left(\Phi,\tilde{T}\right)$ allows a
number of simplifications in equation (\ref{2.14}). In this connection,
we divide the set $\Phi$ into second-class subsets $U$ and $V$
given by \[
U=\left\{ \Phi_{1},\Phi_{2},\Phi_{3},\Phi_{4}\right\} \,,\,\, V=\left\{ \Phi_{5},\Phi_{6},\Phi_{7},\Phi_{8}\right\} \,,\]
 so that it is possible to apply the rule \begin{equation}
\left\{ A,B\right\} _{D\left(\Phi\right)}=\left\{ A,B\right\} _{D\left(U\right)}-\left\{ A,V_{a}\right\} _{D\left(U\right)}C^{ab}\left\{ V_{b},B\right\} _{D\left(U\right)}\,,\;\; C^{ac}\left\{ V_{c},V_{b}\right\} _{D\left(U\right)}=\delta_{b}^{a}\,,\label{2.15}\end{equation}
 for any dynamical variables $A$ and $B$. As a result, thanks to
the vanishing of $\left\{ V_{b},\epsilon\right\} _{D\left(U\right)}$
on the constraint surface, equation (\ref{2.14}) is simplified to
\begin{equation}
\dot{\boldsymbol\eta}=\left\{ \boldsymbol\eta,\Lambda T+\epsilon\right\} _{D\left(U\right)}\,,\;\; U=0,\;\; T=0\,.\label{2.16}\end{equation}

In the same spirit, we further divide the subset $U$ into two sets
$u=\left(\Phi_{3},\Phi_{4}\right)$ and $v=\left(\Phi_{1},\Phi_{2}\right)$.
This time, application of the rule (\ref{2.15}) gives rise to a new
simplification due to the vanishing of $\left\{ \boldsymbol\eta,\epsilon\right\} _{D\left(u\right)}$.
Thus, \begin{equation}
\left\{ \boldsymbol\eta,\epsilon\right\} _{D\left(U\right)}=-\left\{ \boldsymbol\eta,v_{a}\right\} _{D\left(u\right)}c^{ab}\left\{ v_{b},\epsilon\right\} _{D\left(u\right)}=\zeta\left\{ \boldsymbol\eta,\Phi_{1}\right\} _{D\left(u\right)}\,.\label{2.17}\end{equation}
 Since neither the basic variables nor the constraints $u$ involve
the coordinate $x^{0}$, and $p_{0}$ is alone in the constraint $\Phi_{1}$
as an additive factor, we can eliminate $p_{0}$ altogether from (\ref{2.17}).
Consequently, we are now able to take $\Phi_{2}\equiv0$ identically,
and thus substitute $x^{0}=\zeta\tau$ into the same brackets. Finally,
we express $\pi_{k}$ in terms of $\xi^{k}$ in $\tilde{\omega}$,
using the constraints $u$, so that the equations of motion in the
reduced phase space space (\ref{2.13}) become \begin{equation}
\dot{\boldsymbol\eta}=\left\{ \boldsymbol\eta,\mathcal{H}_{\mathrm{eff}}+\Lambda T\right\} _{D\left(u\right)},\;\;\; u_{k}=\pi_{k}+i\xi_{k}=0,\;\;\; T=0,\label{2.18}\end{equation}
 where $\mathcal{H}_{\mathrm{eff}}$ is an effective Hamiltonian given
by \begin{align}
 & \mathcal{H}_{\mathrm{eff}}=\left[\zeta qA_{0}+\omega\right]_{x^{0}=\zeta\tau}\,,\;\;\omega=\left.\tilde{\omega}\right|_{\pi_{k}=-i\xi^{k}}=\sqrt{\omega_{0}^{2}+\rho}\,,\nonumber \\
 & \omega_{0}=\sqrt{m^{2}+\left(p_{k}+qA_{k}\right)^{2}-2iqF_{kl}\xi^{k}\xi^{l}}\,,\;\;\rho=\frac{-4i\zeta qF_{k0}}{\omega_{0}+m}\left(p_{l}+qA_{l}\right)\xi^{k}\xi^{l},\label{2.19}\end{align}
 and $T$ (\ref{2.12}) is an effective first-class constraint.

The nonzero Dirac brackets between the basic variables $\boldsymbol\eta$
are given by \begin{equation}
\left\{ x^{k},p_{l}\right\} _{D\left(u\right)}=\delta_{l}^{k}\,,\,\,\left\{ \xi_{k},\xi_{l}\right\} _{D\left(u\right)}=-\left\{ \pi_{k},\pi_{l}\right\} _{D\left(u\right)}=i\left\{ \xi_{k},\pi_{l}\right\} _{D\left(u\right)}=\frac{i}{2}\eta_{kl}\,.\label{2.20}\end{equation}

\section{Quantization}

\subsection{Operators of basic variables}

The equal-time commutation relations for the operators $\hat{X}^{k}$,
$\hat{P}_{k}$, $\hat{\Xi}^{k}$, and $\hat{\zeta}$, corresponding
to the basic variables $x^{k}$, $p_{k}$, $\xi^{k}$, and $\zeta$,
are defined according to their Dirac brackets (\ref{2.20}). The nonzero
comutators $\left[\,,\,\right]$ (anticomutators $\left[\,,\,\right]_{+}$)
are \begin{equation}
\left[\hat{X}^{k},\hat{P}_{l}\right]=i\hbar\delta_{l}^{k}\,,\;\;\;[\hat{\Xi}^{k},\hat{\Xi}^{l}]_{+}=-\frac{\hbar}{2}\eta^{kl}\,.\label{3.1}\end{equation}
 We assume $\hat{\zeta}^{2}=1$ and select a preliminary state space
$\mathcal{R}$ of $\mathbf{x}$-dependent $16$-component columns
$\underline{\boldsymbol\Psi}(\mathbf{x})$, \begin{equation}
\underline{\boldsymbol\Psi}\left(\mathbf{x}\right)=\left(\begin{array}{c}
\Psi_{+1}\left(\mathbf{x}\right)\\
\Psi_{-1}\left(\mathbf{x}\right)\end{array}\right),\label{3.2}\end{equation}
 where $\Psi_{\zeta}(\mathbf{x})$, $\zeta=\pm1$, are $8$-component
columns. The inner product in $\mathcal{R}$ is defined as follows%
\footnote{In what follows, we define the bilinear form $\left(\psi,\varphi\right)$
as \[
\left(\psi,\varphi\right)=\int\psi^{\dagger}\varphi d\mathbf{x}\]
 for vectors of any finite number of components.%
}, \begin{equation}
\left(\underline{\boldsymbol\Psi}|\underline{\boldsymbol\Psi}^{\prime}\right)=\left(\Psi_{+1},\Psi_{+1}^{\prime}\right)+\left(\Psi_{-1}^{\prime},\Psi_{-1}\right)\,,\;\;\;\left(\Psi,\Psi^{\prime}\right)=\int\Psi^{\dagger}(\mathbf{x})\Psi^{\prime}(\mathbf{x})d\mathbf{x}\,.\label{3.3}\end{equation}
 Later on, we shall see this construction of the inner product is
Lorentz-invariant.

We realize all the operators in the following block-diagonal form%
\footnote{Here and in what follows we use the notation $\mathrm{bdiag}\left(A,B\right)=\left(\begin{array}{cc}
A & 0\\
0 & B\end{array}\right)$, where $A$ and $B$ are matrices.%
}, \begin{align}
 & \hat{X}^{k}=x^{k}I_{16}\,,\;\;\;\hat{P}_{k}=\hat{p}_{k}I_{16}\,,\;\;\;\hat{p}_{k}=-i\hbar\partial_{k}\,,\nonumber \\
 & \hat{\Xi}^{k}=\mathrm{bdiag}\left(\hat{\xi}^{k},\hat{\xi}^{k}\right),\;\;\hat{\zeta}=\mathrm{bdiag}\left(I_{8},-I_{8}\right)\,.\label{3.4}\end{align}
 Here, $I_{16}$ and $I_{8}$ are the $16\times16$ and $8\times8$
unit matrices, respectively, whereas $\hat{\xi}^{k}$ are $8\times8$
matrices which obey the equal-time commutation relations \[
\left[\hat{\xi}^{k},\hat{\xi}^{l}\right]_{+}=-\frac{\hbar}{2}\eta^{kl}.\]

\subsection{Hamiltonian, first-class constraint and spin variables}

Let us construct the operator which is a quantum version of the classical
function $\mathcal{H}_{\mathrm{eff}}$ (\ref{2.19}). We select it
as follows \begin{equation}
\mathcal{H}_{\mathrm{eff}}\rightarrow\underline{\hat{H}}=q\hat{\zeta}\hat{A}_{0}+\underline{\hat{\Omega}}=\mathrm{bdiag}\left(\hat{H}_{+1},\hat{H}_{-1}\right),\label{3.5}\end{equation}
 where $\hat{A}_{0}=\mathrm{bdiag}\left(\left.A_{0}\right|_{x^{0}=\tau}I_{8},\,\left.A_{0}\right|_{x^{0}=-\tau}I_{8}\right)$
, $\underline{\hat{\Omega}}=\mathrm{bdiag}\left(\hat{\Omega}_{+1},\hat{\Omega}_{-1}\right)$,
$\;\hat{\Omega}_{\zeta}=\left.\hat{\omega}_{0}\right|_{x^{0}=\zeta\tau},$
and \begin{equation}
\hat{H}_{\zeta}=q\zeta\left.A_{0}\right|_{x^{0}=\zeta\tau}I_{8}+\hat{\Omega}_{\zeta}=\left.\left(\zeta qA_{0}I_{8}+\hat{\omega}_{0}\right)\right|_{x^{0}=\zeta\tau}\,.\label{3.6}\end{equation}

We realize the operator $\underline{\hat{T}}$ corresponding to the
first-class constraint $T$ (\ref{2.12}) by the prescription $\underline{\hat{T}}=\left.T\right|_{\zeta=\hat{\zeta},\xi=\hat{\Xi}}$
and the classical bifermionic constant $\theta$ as a constant matrix
$\hat{\Theta}=\mathrm{bdiag}\left(\hat{\theta},\hat{\theta}\right)$,
where $\hat{\theta}$ are $8\times8$ matrices. Thus, $\underline{\hat{T}}$
has a block-diagonal structure, \begin{equation}
\underline{\hat{T}}=m\left(2i\hat{\zeta}\hat{\Xi}^{1}\hat{\Xi}^{2}+\frac{1}{2}\hat{\Theta}\right)=\mathrm{bdiag}\left(\hat{t}_{1},\hat{t}_{-1}\right)\,,\;\;\hat{t}_{\zeta}=m\left(2i\zeta\hat{\xi}^{1}\hat{\xi}^{2}+\frac{1}{2}\hat{\theta}\right)\,.\label{3.7}\end{equation}
 We are going to use this operator to fix the gauge at the quantum
level according to Dirac, $\underline{\boldsymbol\Psi}\in\mathcal{R}$,
$\underline{\hat{T}}\,\underline{\boldsymbol\Psi}=0$. One can see
this condition implies \begin{equation}
\hat{t}_{\zeta}\Psi_{\zeta}=0.\label{3.8}\end{equation}
 The conservation of (\ref{3.8}) in time is guaranteed by $\left[\underline{\hat{T}},\underline{\hat{H}}\right]=0$,
which follows from the corresponding relation of the classical theory,
$\left\{ T,\mathcal{H}_{\mathrm{eff}}\right\} _{D\left(u\right)}=0$.
It is equivalent to \begin{equation}
\left[\hat{t}_{\zeta},\hat{\Omega}_{\zeta}\right]=0.\label{3.9}\end{equation}

Now we postulate a manifest form for the operator $\hat{\Omega}_{\zeta}$,
subject to the relation (\ref{3.9}). This form ensures the hermiticity
of the operator $\underline{\hat{H}}$ with respect to the inner product
(\ref{3.3}), provides the gauge invariance under $U(1)$ transformations,
and Lorentz invariance of the inner product (\ref{3.3}). Here, we
note that $\hat{\omega}_{0}$ cannot be realized in analogy with the
$3+1$ dimensional case \cite{GavGi00b,GavGi01}, that is, in the
form of the familiar one-particle Dirac Hamiltonian, and with $\hat{\xi}^{k}$
proportional to gamma-matrices. This is because $\hat{t}_{\zeta}$
contains the term $\hat{\xi}^{1}\hat{\xi}^{2}$, which would not commute
with $\hat{\omega}_{0}$ in such a realization. In order to fulfill
the condition (\ref{3.9}), and simultaneously ensure that $\hat{\omega}_{0}^{2}$
corresponds to the classical $\omega_{0}^{2}$, we select \begin{equation}
\hat{\omega}_{0}=\left(\begin{array}{cc}
0 & m-\gamma^{k}\left(\hat{p}_{k}+qA_{k}\right)\\
m+\gamma^{k}\left(\hat{p}_{k}+qA_{k}\right) & 0\end{array}\right),\label{3.10}\end{equation}
 where $\gamma^{k},$ $k=1,2$, are any $4\times4$ matrices that
obey the relation $\left[\gamma^{k},\gamma^{l}\right]_{+}=-2\delta_{kl}$.
In fact, we can consider them as two matrices of a specific $4\times4$
realization of gamma-matrices in $2+1$ dimensions, $\left[\gamma^{\mu},\gamma^{\nu}\right]_{+}=2\eta^{\mu\nu}$.
Moreover, we can consider these matrices as a part of the complete
set of gamma-matrices in $3+1$ dimensions, for which it is convenient
to select the following representation \cite{AMW89}\begin{equation}
\gamma^{0}=\left(\begin{array}{cc}
\sigma^{3} & 0\\
0 & -\sigma^{3}\end{array}\right),\;\;\gamma^{1}=\left(\begin{array}{cc}
i\sigma^{2} & 0\\
0 & -i\sigma^{2}\end{array}\right),\;\;\gamma^{2}=\left(\begin{array}{cc}
-i\sigma^{1} & 0\\
0 & i\sigma^{1}\end{array}\right),\;\;\gamma^{3}=\left(\begin{array}{cc}
0 & I_{2}\\
-I_{2} & 0\end{array}\right),\label{3.14}\end{equation}
 where $\sigma^{k}$ are the Pauli matrices, and $I_{2}$ is the $2\times2$
unit matrix.

Then, the expression $\underline{\hat{\Omega}}^{2}=\left.\mathrm{bdiag}\left(\hat{\omega}_{0}^{2},\hat{\omega}_{0}^{2}\right)\right|_{x^{0}=\zeta\tau}$
, where \begin{equation}
\hat{\omega}_{0}^{2}=\left(\begin{array}{cc}
m^{2}+\left(\hat{p}_{k}+qA_{k}\right)^{2}+\frac{i\hbar}{2}qF_{kl}\gamma^{k}\gamma^{l} & 0\\
0 & m^{2}+\left(\hat{p}_{k}+qA_{k}\right)^{2}+\frac{i\hbar}{2}qF_{kl}\gamma^{k}\gamma^{l}\end{array}\right),\label{3.11}\end{equation}
 is consistent with the semiclassical limit (see Appendix).

The realization of the operators $\hat{\xi}^{k}$ , $k=1,2$, and
of the matrix $\hat{\theta}$, corresponding to the classical quantity
$\theta$, is constrained by the relation (\ref{3.9}). The latter
relation is a constraint on $\hat{t}_{\zeta}$, and implies that $\hat{\xi}^{1}\hat{\xi}^{2}$
and $\hat{\theta}$ must commute with $\hat{\omega}_{0}$. Additionally,
we require the condition $\left[\hat{\xi}^{k},\hat{\theta}\right]=0$
in accordance with the classical theory.

The above restraints are not sufficient to single out a representation,
so we impose further restrictions to the form of $\hat{\xi}^{1}\hat{\xi}^{2}$
and $\hat{\theta}$. The matrix of $\hat{\xi}^{1}\hat{\xi}^{2}$ is
chosen to be composed of blocks which are products of two $4\times4$
gamma-matrices, and the matrix $\hat{\theta}$ is chosen to be diagonal
with eigenvalues $\pm\hbar$. The last restriction is consistent with
the relation $\hat{\theta}^{2}=\hbar^{2}$, valid in the subspace
of states satisfying the condition (\ref{3.8}), where $\hat{\theta}^{2}$
can be identified with $\left(4i\hat{\xi}^{1}\hat{\xi}^{2}\right)^{2}=\hbar^{2}$.
Moreover, it is clear that $\hat{\xi}^{1}\hat{\xi}^{2}$ cannot be
the unit matrix, since this would lead to a contradiction with the
commutation relations for $\hat{\xi}^{k}$. There is only one realization
in the space of $8\times8$ matrices which fulfills all the aforementioned
demands, viz., \begin{equation}
\hat{\theta}=\hbar\left(\begin{array}{cc}
\gamma^{0}\Sigma^{3} & 0\\
0 & \gamma^{0}\Sigma^{3}\end{array}\right),\;\;\;\hat{\xi}^{1}\hat{\xi}^{2}=\frac{i\hbar}{4}\left(\begin{array}{cc}
0 & \Sigma^{3}\\
\Sigma^{3} & 0\end{array}\right)\,,\label{3.12}\end{equation}
 where $\Sigma^{3}=i\gamma^{1}\gamma^{2}$. Then, \begin{equation}
\hat{\xi}^{1}=\frac{i}{2}\hbar^{1/2}\left(\begin{array}{cc}
0 & \gamma^{1}\\
\gamma^{1} & 0\end{array}\right),\;\;\;\hat{\xi}^{2}=\frac{i}{2}\hbar^{1/2}\left(\begin{array}{cc}
\gamma^{2} & 0\\
0 & \gamma^{2}\end{array}\right).\label{3.13}\end{equation}

Taking into account the concrete realization of the operator $\hat{t}_{\zeta}$,
we can see that states $\Psi_{\zeta}$ that obey the condition (\ref{3.8})
have the following form \begin{equation}
\Psi_{\zeta}(\tau,\mathbf{x})=\frac{1}{\sqrt{2}}\left(\begin{array}{c}
\boldsymbol\psi_{\zeta}(\tau,\mathbf{x})\\
\zeta\gamma^{0}\boldsymbol\psi_{\zeta}(\tau,\mathbf{x})\end{array}\right)\,,\label{3.15}\end{equation}
 where the factor $1/\sqrt{2}$ has been inserted for convenience.

\subsection{Schrödinger equation}

The Schrödinger equation \begin{equation}
i\hbar\partial_{\tau}\underline{\boldsymbol\Psi}=\left(\underline{\hat{H}}+\Lambda\underline{\hat{T}}\right)\underline{\boldsymbol\Psi},\label{3.16}\end{equation}
 with $\underline{\hat{H}}$ given by (\ref{3.5}), for vectors $\underline{\boldsymbol\Psi}$
subject to $\underline{\hat{T}}\,\underline{\boldsymbol\Psi}=0$,
has the form \begin{equation}
i\hbar\partial_{\tau}\underline{\boldsymbol\Psi}=\underline{\hat{H}}\,\underline{\boldsymbol\Psi}.\label{3.17}\end{equation}
 Solutions of the above equation can be chosen as eigenstates of the
matrix $\hat{\Theta}=\mathrm{bdiag}\left(\hat{\theta},\hat{\theta}\right)$.
Let us denote eigenstates of $\hat{\theta}$ by $\Psi_{\zeta,\theta}$,
which are subject to $\hat{\theta}\Psi_{\zeta,\theta}=\theta\hbar\Psi_{\zeta,\theta}$,
with the eigenvalues $\theta=\pm1$. The latter implies that these
solutions have the specific structure \begin{align}
 & \Psi_{\zeta,\theta}(\tau,\mathbf{x})=\frac{1}{\sqrt{2}}\left(\begin{array}{c}
\boldsymbol\psi_{\zeta,\theta}(\tau,\mathbf{x})\\
\zeta\gamma^{0}\boldsymbol\psi_{\zeta,\theta}(\tau,\mathbf{x})\end{array}\right),\nonumber \\
 & \boldsymbol\psi_{\zeta,+1}(\tau,\mathbf{x})=\left(\begin{array}{c}
\psi_{\zeta}^{(+1)}(\tau,\mathbf{x})\\
0\end{array}\right),\;\;\boldsymbol\psi_{\zeta,-1}(\tau,\mathbf{x})=\left(\begin{array}{c}
0\\
\sigma^{1}\psi_{\zeta}^{(-1)}(\tau,\mathbf{x})\end{array}\right),\label{3.18}\end{align}
 where $\psi_{\zeta}^{(\theta)}(\tau,\mathbf{x})$ are $2$-component
columns. We can see that, due to the constraint (\ref{3.8}), these
states obey the eigenvalue equation $-2i\hat{\xi}^{1}\hat{\xi}^{2}\Psi_{\zeta,\theta}=\frac{\hbar}{2}\theta\zeta\Psi_{\zeta,\theta}\,$.
Consequently, the eigenvalues $\theta$ label different particle species.
Therefore, at the quantum level, we observe the noninvariance of the
theory with respect to reflections of one of the coordinate axes.

In components, and in terms of the physical time $x^{0}=\zeta\tau$,
the Schrödinger equation (\ref{3.17}) implies \begin{equation}
\left[\gamma^{\mu}\left(i\hbar\partial_{\mu}-qA_{\mu}\right)-m\right]\boldsymbol\psi_{\zeta,\theta}\left(\zeta x^{0},\mathbf{x}\right)=0.\label{3.19}\end{equation}

By analogy with the $3+1$ case, one can regard $\hat{\zeta}$ as
the charge-sign operator. Let us consider the states $\underline{\boldsymbol\Psi}_{\zeta}$
with a definite charge $q\zeta$. These states satisfy the eigenvalue
equation $\hat{\zeta}\underline{\boldsymbol\Psi}_{\zeta}=\zeta\underline{\boldsymbol\Psi}_{\zeta}$.
Therefore, states with definite charge $\pm q$ are represented by
(\ref{3.2}) with $\Psi_{\mp1}=0$. The wave functions $\Psi_{\pm1}$
are parameterized by the physical time $\tau=\pm x^{0}$.

It is clear that (\ref{3.19}) for $\zeta=+1$ is the Dirac equation
in the four-spinor representation (lacking the third spatial coordinate)
for a particle with charge $+q$. The solutions of (\ref{3.19}) with
$\zeta=-1$ can be brought into correspondence with the Dirac equation
in the four-spinor representation for a particle $\boldsymbol\psi^{c}\left(x^{0}\right)$
with charge $-q$, by the rule $\boldsymbol\psi^{c}\left(x^{0}\right)=\gamma^{2}\boldsymbol\psi_{-1}^{\ast}\left(-x^{0}\right)$.
In order to arrive at the two-spinor representation of the $2+1$
Dirac equations for particles with charge $\pm q$, we shall use the
decomposition (\ref{3.18}) of the four-component column $\boldsymbol\psi_{\zeta,\theta}$
into two-component columns $\psi_{\zeta}^{(\theta)}$. For $\zeta=+1$,
the equation (\ref{3.19}) decomposes into \begin{equation}
\left[\Gamma_{\theta}^{\mu}\left(i\hbar\partial_{\mu}-qA_{\mu}\right)-m\right]\psi^{\left(\theta\right)}\left(x\right)=0,\;\;\psi^{\left(\theta\right)}\left(x\right)\equiv\psi_{+1}^{\left(\theta\right)}\left(x^{0},\mathbf{x}\right),\label{3.20}\end{equation}
 where $\Gamma_{\theta}^{\mu}$ are two inequivalent sets of gamma-matrices
in $2+1$ dimensions, given by\begin{equation}
\Gamma_{+1}^{0}=\Gamma_{-1}^{0}=\sigma^{3},\Gamma_{+1}^{1}=\Gamma_{-1}^{1}=i\sigma^{2},\Gamma_{+1}^{2}=-\Gamma_{-1}^{2}=-i\sigma^{1},\label{a2}\end{equation}
 where $\sigma^{i}$ are the Pauli matrices. Unless otherwise specified,
we shall always assume $\Gamma=\Gamma_{\theta}$, $\psi=\psi^{(\theta)}$,
$\bar{\psi}=\bar{\psi}^{(\theta)}$, and so on. The analogous equation
for $\zeta=-1$ has the form \begin{equation}
\left[\Gamma_{\theta}^{\mu}\left(i\hbar\partial_{\mu}+qA_{\mu}\right)-m\right]\psi^{\left(\theta\right)c}\left(x\right)=0,\;\;\;\psi^{\left(\theta\right)c}\left(x\right)\equiv\Gamma_{\theta}^{2}\psi_{-1}^{\left(\theta\right)\ast}\left(-x^{0},\mathbf{x}\right),\label{3.21}\end{equation}
 which is the $2+1$ Dirac equation in the two-spinor representation
for a particle with charge $-q$.

Next, we define the $x^{0}$-representation of states with definite
$\theta$-eigenvalue in terms of the physical time $x^{0}$, \begin{align}
 & \boldsymbol\Psi_{\theta}\left(x\right)=\left(\begin{array}{c}
\Psi_{\theta}\left(x\right)\\
\Psi_{\theta}^{c}\left(x\right)\end{array}\right),\;\;\Psi_{\theta}\left(x\right)=\frac{1}{\sqrt{2}}\left(\begin{array}{c}
\boldsymbol\psi_{\theta}\left(x\right)\\
\gamma^{0}\boldsymbol\psi_{\theta}\left(x\right)\end{array}\right),\;\;\Psi_{\theta}^{c}\left(x\right)=\frac{1}{\sqrt{2}}\left(\begin{array}{c}
\boldsymbol\psi_{\theta}^{c}\left(x\right)\\
\gamma^{0}\boldsymbol\psi_{\theta}^{c}\left(x\right)\end{array}\right),\nonumber \\
 & \boldsymbol\psi_{+1}\left(x\right)=\left(\begin{array}{c}
\psi^{\left(+1\right)}\left(x\right)\\
0\end{array}\right),\;\;\boldsymbol\psi_{-1}\left(x\right)=\left(\begin{array}{c}
0\\
\sigma^{1}\psi^{\left(-1\right)}\left(x\right)\end{array}\right),\nonumber \\
 & \boldsymbol\psi_{+1}^{c}\left(x\right)=\left(\begin{array}{c}
\psi^{\left(+1\right)c}\left(x\right)\\
0\end{array}\right),\;\;\boldsymbol\psi_{-1}^{c}\left(x\right)=\left(\begin{array}{c}
0\\
\sigma^{1}\psi^{\left(-1\right)c}\left(x\right)\end{array}\right).\label{3.22}\end{align}
 The charge-conjugate components $\Psi_{\theta}^{c}\left(x\right)$
have been obtained by the rule \[
\Psi_{\theta}^{c}\left(x\right)=\left(\begin{array}{cc}
0 & \gamma^{0}\gamma^{2}\\
\gamma^{0}\gamma^{2} & 0\end{array}\right)\Psi_{-1,\theta}^{\ast}\left(-x^{0},\mathbf{x}\right)\,,\]
 and the inner product (\ref{3.3}) in the $x^{0}$-representation
is reformulated as \begin{equation}
\left(\boldsymbol\Psi_{\theta},\boldsymbol\Psi_{\theta}^{\prime}\right)=\left(\Psi_{\theta},\Psi_{\theta}^{\prime}\right)+\left(\Psi_{\theta}^{c},\Psi_{\theta}^{c\prime}\right)\,,\label{3.23}\end{equation}
 The states $\boldsymbol\Psi_{\theta}\left(x\right)$ satisfy the
evolution equation \begin{equation}
i\hbar\partial_{0}\boldsymbol\Psi_{\theta}\left(x^{0},\mathbf{x}\right)=\mathbf{\hat{H}}\Psi_{\theta}\left(x^{0},\mathbf{x}\right),\;\;\mathbf{\hat{H}}=\mathrm{bdiag}\left(\hat{H}\left(x^{0}\right),\hat{H}^{c}\left(x^{0}\right)\right),\label{3.24}\end{equation}
 where \begin{align}
 & \hat{H}\left(x^{0}\right)=qA_{0}I_{8}+\hat{\omega}_{0}\,,\;\;\;\hat{H}^{c}\left(x^{0}\right)=\left.\hat{H}\left(x^{0}\right)\right|_{q\rightarrow-q}=-qA_{0}I_{8}+\hat{\omega}_{0}^{c}\,,\nonumber \\
 & \hat{\omega}_{0}^{c}=\left.\hat{\omega}_{0}^{c}\right|_{q\rightarrow-q}=\left(\begin{array}{cc}
0 & \gamma^{0}\gamma^{2}\\
\gamma^{0}\gamma^{2} & 0\end{array}\right)\hat{\omega}_{0}^{\ast}\left(\begin{array}{cc}
0 & \gamma^{0}\gamma^{2}\\
\gamma^{0}\gamma^{2} & 0\end{array}\right).\label{3.25}\end{align}
 The operator $\underline{\hat{\Omega}}$ in the $x^{0}$-representation
has the form $\hat{\Omega}=\mathrm{bdiag}\left(\hat{\omega}_{0},\hat{\omega}_{0}^{c}\right)$,
while the operator $\underline{\hat{T}}$ in such a representation
reads \begin{equation}
\hat{T}=\mathrm{bdiag}\left(\hat{t},\hat{t}\right)\,,\,\,\hat{t}=2i\hat{\xi}^{1}\hat{\xi}^{2}+\frac{1}{2}\hat{\theta}\,.\label{3.26}\end{equation}
 Then, states in the $x^{0}$-representation satisfy the condition
\begin{equation}
\hat{T}\boldsymbol\Psi_{\theta}=0\,.\label{3.27}\end{equation}
 We can also see that the inner product (\ref{3.23}) reduces to the
standard inner product between $2+1$ spinors \begin{equation}
\left(\boldsymbol\Psi_{\theta},\boldsymbol\Psi_{\theta}^{\prime}\right)=\left(\psi^{\left(\theta\right)},\psi^{\left(\theta\right)\prime}\right)+\left(\psi^{\left(\theta\right)c},\psi^{\left(\theta\right)c\prime}\right).\label{3.28}\end{equation}
 Thus, in terms of $\psi^{(\theta)}$, treated as Dirac spinors, the
inner product (\ref{3.23}) is Lorentz-invariant.

We note that if one abandons condition (\ref{3.27}), one gets a $P$-invariant
QM, which can be obtained by dimensional reduction of the $3+1$ QM
given by \cite{GavGi00b}. In the event that (\ref{3.27}) is no longer
valid, states with distinct $\theta$-values are allowed to interfere.

We have considered a realization in which both species of particles
are described in the same Hilbert space. As we shall see in the section
4, this fact provides some advantages in studying questions related
to spin polarization. For this reason, we have introduced in (\ref{3.2})
the space $\mathcal{R}$ of $\mathbf{x}$-dependent $16$-component
columns $\underline{\boldsymbol\Psi}(\mathbf{x})$. However, if one
assumes a given value of the parameter $\theta$ fixed from the beginning,
then one can obtain a physically equivalent realization of the Hilbert
space of vectors $\Psi_{\zeta}(\mathbf{x})$ in (\ref{3.2}) as $4$-component
columns. Accordingly, one should use the $2\times2$ gamma-matrices,
instead of the $4\times4$ gamma-matrices, in the representations
(\ref{3.10}) and (\ref{3.13}) for the operators $\hat{\omega}_{0}$
and $\hat{\xi}^{k}$. In such a realization, one immediately arrives
at the above-described two-spinor representation of Dirac spinors,
thus avoiding the intermediate description in terms of four-spinors.

\subsection{Physical sector}

The preliminary state space contains an infinite number of negative-energy
states. We restrict it to a physical subspace $\mathcal{R}_{\mathrm{ph}}$
where these negative-energy states are absent. Namely, taking into
account the correspondence principle, we demand that the operator
$\hat{\Omega}$ be positive definite in $\mathcal{R}_{\mathrm{ph}}$.

In further considerations we use a time-independent background, which
is enough for our purposes. In time-independent backgrounds the Dirac
equation is reduced to its stationary form \begin{equation}
\hat{h}\psi\left(\mathbf{x}\right)=\varepsilon\psi\left(\mathbf{x}\right),\;\;\psi\left(x\right)=\exp\left(-i\varepsilon x^{0}\right)\psi\left(\mathbf{x}\right)\,,\label{a8}\end{equation}
 where \begin{equation}
\hat{h}=-\Gamma^{0}\Gamma^{k}P_{k}+\Gamma^{0}m+qA_{0}\,,\; P_{k}=i\partial_{k}-qA_{k}\,.\label{a4}\end{equation}
 We square this equation with the ansatz \begin{equation}
\psi\left(\mathbf{x}\right)=\left[\Gamma^{0}\left(\varepsilon-qA_{0}\right)+\Gamma^{k}P_{k}+m\right]\varphi\left(\mathbf{x}\right)\,.\label{a9}\end{equation}
 Then $\varphi\left(\mathbf{x}\right)$ satisfies the equation \begin{equation}
\left[\left(\varepsilon-qA_{0}\right)^{2}-D\right]\varphi\left(\mathbf{x}\right)=0,\label{a10}\end{equation}
 where $D=m^{2}-P_{k}P^{k}+\frac{i}{4}qF_{\mu\nu}\left[\Gamma^{\mu},\Gamma^{\nu}\right]$.
A pair $\left(\varepsilon,\varphi\right)$ is a solution of the above
equation if it obeys either \begin{equation}
\varepsilon=qA_{0}+\sqrt{\varphi^{-1}D\varphi}\Rightarrow\varepsilon-qA_{0}>0,\label{a11}\end{equation}
 or \begin{equation}
\varepsilon=qA_{0}-\sqrt{\varphi^{-1}D\varphi}\Rightarrow\varepsilon-qA_{0}<0.\label{a12}\end{equation}

Let us denote by $\left(\varepsilon_{+,n},\varphi_{+,n}\right)$ solutions
for positive values of $\varepsilon-qA_{0}$, i.e., those for the
upper branch of the energy spectrum, and by $\left(\varepsilon_{-,\alpha},\varphi_{-,\alpha}\right)$
solutions for negative values of $\varepsilon-qA_{0}$, i.e., those
for the lower branch of the energy spectrum. Here, $n$ and $\alpha$
are quantum numbers which account for a possible asymmetry between
both branches of the energy spectrum.

Solutions $\psi_{+,n}$ and $\psi_{-,\alpha}$ of (\ref{a8}), constructed
from $\varphi_{+,n}$ and $\varphi_{-,\alpha}$, obey the orthogonality
and completeness relations \begin{align}
 & \left(\psi_{+,n},\psi_{+,m}\right)=\delta_{nm}\,,\;\;\left(\psi_{-,\alpha},\psi_{-,\beta}\right)=\delta_{\alpha\beta}\,,\;\;\left(\psi_{+,n},\psi_{-,\alpha}\right)=0,\label{a13}\\
 & \sum_{n,\alpha}\left[\psi_{+,n}\left(x\right)\psi_{+,n}^{\dagger}\left(y\right)+\psi_{-,\alpha}\left(x\right)\psi_{-,\alpha}^{\dagger}\left(y\right)\right]=\delta\left(\mathbf{x}-\mathbf{y}\right),\;\; x_{0}=y_{0},\label{a14}\end{align}
 where \begin{equation}
\psi_{+,n}\left(x\right)=e^{-i\varepsilon_{+,n}x^{0}}\psi_{+,n}\left(\mathbf{x}\right),\;\;\psi_{-,\alpha}\left(x\right)=e^{-i\varepsilon_{-,\alpha}x^{0}}\psi_{-,\alpha}\left(\mathbf{x}\right)\,.\label{a15}\end{equation}

A solution of the eigenvalue problem $\hat{h}^{c}\psi^{c}\left(\mathbf{x}\right)=\varepsilon^{c}\psi^{c}\left(\mathbf{x}\right)$
for the charge-conjugated Hamiltonian, $\hat{h}^{c}=\left.\hat{h}\left(q\right)\right|_{q\rightarrow-q}$,
can be analyzed in a similar manner, and \begin{equation}
\psi_{+,\alpha}^{c}=\Gamma^{2}\psi_{-,\alpha}^{\ast}\,,\;\;\psi_{-,n}^{c}=\Gamma^{2}\psi_{+,n}^{\ast}\,,\;\;\varepsilon_{+,\alpha}^{c}=-\varepsilon_{-,\alpha}\,,\;\;\varepsilon_{-,n}^{c}=-\varepsilon_{+,n}\,.\label{a16}\end{equation}

The bases vectors of $\mathcal{R}_{\mathrm{ph}}$, at a given instant
of time, have the general structure (\ref{3.22}), with the two-spinors
being the solutions $\psi_{+,n}^{\left(\theta\right)}$ (\ref{a15})
and $\psi_{+,\alpha}^{\left(\theta\right)c}$ (\ref{a16}) of the
stationary Dirac equation, corresponding to the energy eigenvalues
$\epsilon_{+,n}^{\left(\theta\right)}$ and $\epsilon_{+,\alpha}^{\left(\theta\right)c}$.
In addition, we require that the basis vectors be eigenvectors of
the charge operator $q\hat{\zeta}$, exactly as is done in QFT when
imposing the superselection rule (\ref{4.1}). Thus, the basis vectors
\begin{equation}
\boldsymbol\Psi_{\theta,+,n}=\left(\begin{array}{c}
\Psi_{\theta,+,n}\\
0\end{array}\right),\;\;\boldsymbol\Psi_{\theta,+,\alpha}^{c}=\left(\begin{array}{c}
0\\
\Psi_{\theta,+,\alpha}^{c}\end{array}\right)\,,\label{3.29}\end{equation}
 are eigenvectors of the charge operator $q\hat{\zeta}$\[
q\hat{\zeta}\boldsymbol\Psi_{\theta,+,n}=q\boldsymbol\Psi_{\theta,+,n}\,,\;\; q\hat{\zeta}\boldsymbol\Psi_{\theta,+,\alpha}^{c}=-q\boldsymbol\Psi_{\theta,+,\alpha}^{c}\,,\]
 and of the Hamiltonian operator (\ref{3.24}) \[
\mathbf{\hat{H}}\boldsymbol\Psi_{\theta,+,n}=\epsilon_{+,n}^{\left(\theta\right)}\boldsymbol\Psi_{\theta,+,n},\;\mathbf{\hat{H}}\boldsymbol\Psi_{\theta,+,n}^{c}=\epsilon_{+,n}^{\left(\theta\right)c}\boldsymbol\Psi_{\theta,+,n}^{c},\]
 The positivity of $\hat{\Omega}$ can be easily established with
respect to the basis vectors $\boldsymbol\Psi_{\theta,+,n}$ and $\boldsymbol\Psi_{\theta,+,n}^{c}$,
\begin{align}
 & \left(\boldsymbol\Psi_{\theta,+,n}\,,\hat{\Omega}\boldsymbol\Psi_{\theta,+,n}\right)=\left(\psi_{+,n}^{(\theta)},\left(\epsilon_{+,n}^{(\theta)}-qA_{0}\right)\psi_{+,n}^{(\theta)}\right)>0\,,\nonumber \\
 & \left(\boldsymbol\Psi_{\theta,+,n}^{c}\,,\hat{\Omega}\boldsymbol\Psi_{\theta,+,n}^{c}\right)=\left(\psi_{+,n}^{(\theta)c},\left(\epsilon_{+,n}^{(\theta)c}+qA_{0}\right)\psi_{+,n}^{(\theta)c}\right)>0\,.\label{3.29b}\end{align}

Thus, the physical subspace $\mathcal{R}_{\mathrm{ph}}$ is formed
by the vectors of the form $\boldsymbol\Psi_{\theta,+}$ and $\boldsymbol\Psi_{\theta,+}^{c}\,,$
which are linear combinations of the states $\boldsymbol\Psi_{\theta,+,n}$
and $\boldsymbol\Psi_{\theta,+,n}^{c}$ respectively. The vectors
from $\mathcal{R}_{\mathrm{ph}}$ satisfy the Schrödinger equation
with the Hamiltonian (\ref{3.24}). The inner product (\ref{3.23})
between charged states from $\mathcal{R}_{\mathrm{ph}}$ of the same
sign is given by \begin{equation}
\left(\boldsymbol\Psi_{\theta,+}\,,\boldsymbol\Psi_{\theta,+}^{\prime}\right)=\left(\Psi_{\theta,+}\;,\Psi_{\theta,+}^{\prime}\right)\,,\;\;\;\left(\boldsymbol\Psi_{\theta,+}^{c}\,,\boldsymbol\Psi_{\theta,+}^{c\prime}\right)=\left(\Psi_{\theta,+}^{c}\;,\Psi_{\theta,+}^{c\prime}\right)\,,\label{3.30}\end{equation}
 and it vanishes between charged states of different sign. And following
(\ref{3.29b}), we see that the operator $\hat{\Omega}$ is positive
definite, \begin{equation}
\left(\boldsymbol\Psi_{\theta,+}\,,\hat{\Omega}\boldsymbol\Psi_{\theta,+}\right)>0,\;\left(\boldsymbol\Psi_{\theta,+}^{c}\,,\hat{\Omega}\boldsymbol\Psi_{\theta,+}^{c}\right)>0\,,\label{3.31}\end{equation}

Let us introduce the conserved spin polarization operator $\hat{S}$
, \[
\,\hat{S}=-2i\hat{\Xi}^{1}\hat{\Xi}^{2}\hat{\zeta}\,.\]
 One can see (taking into account the relation (\ref{3.27})) that
this is a conserved operator, whose eigenvectors are $\boldsymbol\Psi_{\theta}$
and $\boldsymbol\Psi_{\theta}^{c}$, \begin{equation}
\hat{S}\boldsymbol\Psi_{\theta}=\frac{\hbar}{2}\theta\boldsymbol\Psi_{\theta}\,,\,\,\hat{S}\boldsymbol\Psi_{\theta}^{c}=-\frac{\hbar}{2}\theta\boldsymbol\Psi_{\theta}^{c}\,,\label{3.33}\end{equation}
 One can note that the operator $\hat{\zeta}\hat{S}$ acts on the
states $\boldsymbol\Psi_{\theta}$ and $\boldsymbol\Psi_{\theta}^{c}$
as a particle species operator, \[
\hat{\zeta}\hat{S}\boldsymbol\Psi_{\theta}=\hbar\theta\boldsymbol\Psi_{\theta}\,,\,\,\hat{\zeta}\hat{S}\boldsymbol\Psi_{\theta}^{c}=\hbar\theta\boldsymbol\Psi_{\theta}^{c}\,.\]

In the present work, we do not exceed the limits of the one-particle
consideration%
\footnote{However, a generalization to the many-particle theory can be made
on the basis of the constructed one-particle representation, and the
existence of eigenvectors for the position operator $\hat{X}$ of
QFT can be demonstrated. This will be presented elsewhere.%
} within the constructed QM. Thus, it is enough to study the overlaps
of the type (\ref{3.30}), as well as one-particle matrix elements 

\[
\left(\boldsymbol\Psi_{\theta,+}\,,\mathcal{F}\boldsymbol\Psi_{\theta,+}^{\prime}\right)=\left(\Psi_{\theta,+}\;,f\Psi_{\theta,+}^{\prime}\right)\,,\;\;\;\left(\boldsymbol\Psi_{\theta,+}^{c}\,,\mathcal{F}\boldsymbol\Psi_{\theta,+}^{c\prime}\right)=\left(\Psi_{\theta,+}^{c}\;,f^{c}\Psi_{\theta,+}^{c\prime}\right)\]
 for functions of physical operators\[
\mathcal{F}\left(\hat{X}^{k},\hat{P}_{k},\hat{\Xi}^{k},\mathbf{\hat{H}}\right)=\mathrm{bdiag}\left(f\left(x^{k},\hat{p}_{k},\hat{\xi}^{k},\hat{H}\left(x^{0}\right)\right),f^{c}\left(x^{k},\hat{p}_{k},\hat{\xi}^{k},\hat{H}^{c}\left(x^{0}\right)\right)\right)\,.\]
Since all matrix elements of odd-component products of the operators
$\hat{\Xi}^{k}$ are zero and the dynamics of the product of two operators
$\hat{\Xi}^{k}$ is trivial due to (\ref{3.33}), it is possible to
reduce the physical subspace $\mathcal{R}_{\mathrm{ph}}$ of QM to
an effective physical subspace required to calculate matrix elements
of the functions of operators $\mathcal{\tilde{F}}\left(\hat{X}^{k},\hat{P}_{k},\mathbf{\hat{H}}\right)$.
Using the decomposition of the $16$-component columns $\boldsymbol\Psi_{\theta}$
and $\boldsymbol\Psi_{\theta}^{c}$ in terms of two-component spinors
$\psi^{\left(\theta\right)}$ and $\psi^{\left(\theta\right)c}$,
we define the effective physical sector $\mathcal{R}_{\mathrm{ph}}^{\mathrm{eff}}$
as the space of states \begin{equation}
\boldsymbol\psi_{+}^{\left(\theta\right)}=\left(\begin{array}{c}
\psi_{+}^{\left(\theta\right)}\\
0\end{array}\right),\;\;\;\boldsymbol\psi_{+}^{\left(\theta\right)c}=\left(\begin{array}{c}
0\\
\psi_{+}^{\left(\theta\right)c}\end{array}\right),\label{3.34}\end{equation}
 where $\psi_{+}^{\left(\theta\right)}$ and $\psi_{+}^{\left(\theta\right)c}$
are the linear envelops of the spinors $\psi_{+,n}^{(\theta)}$ and
$\psi_{+,\alpha}^{(\theta)c}$, respectively. The dynamics in this
representation is governed by the Hamiltonian \begin{equation}
\hat{H}^{\left(\theta\right)}=\mathrm{bdiag}\left(\hat{h}^{\left(\theta\right)},\hat{h}^{\left(\theta\right)c}\right),\mathrm{\;\;}\hat{h}^{\left(\theta\right)}=qA_{0}+\Gamma_{\theta}^{0}\left[m+\Gamma_{\theta}^{k}\left(\hat{p}_{k}+qA_{k}\right)\right],\;\;\hat{h}^{\left(\theta\right)c}=\left.\hat{h}^{\left(\theta\right)}\right|_{q\rightarrow-q}.\label{3.35}\end{equation}
 The operators $\hat{\zeta}$ and $\hat{S}$, which act in the space
$\mathcal{R}_{\mathrm{ph}},$ are reduced to the respective operators
acting in the space $\mathcal{R}_{\mathrm{ph}}^{\mathrm{eff}}$\begin{equation}
\hat{\zeta}=\mathrm{bdiag}\left(I_{2},-I_{2}\right),\;\hat{S}=\theta\frac{\hbar}{2}\mathrm{bdiag}\left(I_{2},-I_{2}\right)\,.\label{3.36}\end{equation}

We can calculate the matrix elements of $\mathcal{\tilde{F}}\left(\hat{X}^{k},\hat{P}_{k},\mathbf{\hat{H}}\right)$
using its representative \[
\mathrm{bdiag}\left(\tilde{f}\left(x^{k},\hat{p}_{k},\hat{h}_{\theta}\right),\tilde{f}^{c}\left(x^{k},\hat{p}_{k},\hat{\xi}^{k},\hat{h}_{\theta}^{c}\right)\right)\]
 in $\mathcal{R}_{\mathrm{ph}}^{\mathrm{eff}}$ as follows, \[
\left(\boldsymbol\Psi_{\theta,+}\;,\mathcal{\tilde{F}}\boldsymbol\Psi_{\theta,+}^{\prime}\right)=\left(\boldsymbol\psi_{+}^{\left(\theta\right)},\tilde{f}\boldsymbol\psi_{+}^{\prime\left(\theta\right)}\right)\,,\;\;\left(\boldsymbol\Psi_{\theta,+}^{c}\;,\mathcal{\tilde{F}}\boldsymbol\Psi_{\theta,+}^{\prime c}\right)=\left(\boldsymbol\psi_{+}^{\left(\theta\right)c},\tilde{f}^{c}\boldsymbol\psi_{+}^{\prime\left(\theta\right)c}\right).\]

\section{Comparison with one-particle sector of QFT}

We shall now give an interpretation of the constructed QM by making
a comparison with the dynamics of the one-particle sector in the QFT
of the Dirac field in $2+1$ dimensions. To this end, we shall first
demonstrate that the one-particle sector (in case it can be consistently
defined) may be formulated as a consistent relativistic QM. Then,
we shall demonstrate that this one-particle sector may be identified
with the QM constructed in the previous section.

To begin with, we recall that the one-particle sector of QFT (as well
as any sector with a definite particle number) can be defined in an
unique way for every moment of time only in the class of external
backgrounds which do not create particles from the vacuum \cite{BirDa82,GreMuR85,FraGiS91,GriMaM88},
such as stationary magnetic fields and non-critical Coulomb fields.
For this reason, we simplify the present discussion to this class
of backgrounds. A generalization to arbitrary backgrounds which do
not violate vacuum stability is possible.

Let us construct the Hilbert space of one-particle states of a given
species as the disjoint union, $\mathcal{R}_{\mathrm{ph}}^{\mathrm{QFT}}=\mathcal{R}_{10}^{\mathrm{QFT}}\cup\mathcal{R}_{01}^{\mathrm{QFT}}$,
$\mathcal{R}_{10}^{\mathrm{QFT}}\cap\mathcal{R}_{01}^{\mathrm{QFT}}=\left\{ 0\right\} $,
of the particle subspace $\mathcal{R}_{10}^{\mathrm{QFT}}$ and the
antiparticle subspace $\mathcal{R}_{01}^{\mathrm{QFT}}$, \[
\left|\boldsymbol\Psi\right\rangle =\left(\sum_{n}f_{n}a_{n}^{+}\left|0\right\rangle ,\sum_{\alpha}\lambda_{\alpha}b_{\alpha}^{+}\left|0\right\rangle \right)\in\mathcal{R}_{\mathrm{ph}}^{1},\;\;\sum_{n}f_{n}a_{n}^{+}\left|0\right\rangle \in\mathcal{R}_{10}^{\mathrm{QFT}}\,,\;\;\sum_{\alpha}\lambda_{\alpha}b_{\alpha}^{+}\left|0\right\rangle \in\mathcal{R}_{01}^{\mathrm{QFT}}\,,\]
 where $\left(a,a^{+}\right)$ and $\left(b,b^{+}\right)$ are annihilation
and creation operators of particles and antiparticles respectively,
and the vacuum state $\left|0\right\rangle $ is the zero vector of
the annihilation operators: $a_{n}\left|0\right\rangle =b_{\alpha}\left|0\right\rangle =0$
for every $n$ and $\alpha$. The arbitrary coefficients $f_{n}$
and $\lambda_{n}$ are subject to the conditions $\sum_{n}\left|f_{n}\right|^{2}<\infty$
and $\sum_{\alpha}\left|\lambda_{\alpha}\right|^{2}<\infty$. Therefore,
physical states $\left|\boldsymbol\Psi\right\rangle $ belong either
to the particle subspace $\mathcal{R}_{10}^{\mathrm{QFT}}$ or to
the antiparticle subspace $\mathcal{R}_{01}^{\mathrm{QFT}}$, in agreement
with the superselection rule \cite{weinberg}. In other words, physical
states $\left|\boldsymbol\Psi\right\rangle $ are eigenstates of the
charge operator $\hat{Q}^{\mathrm{QFT}}$,\begin{equation}
\hat{Q}^{\mathrm{QFT}}=q\left(\sum_{n}a_{n}^{+}a_{n}-\sum_{\alpha}b_{\alpha}^{+}b_{\alpha}\,\right),\label{a20}\end{equation}
\begin{equation}
\hat{Q}^{\mathrm{QFT}}\left|\boldsymbol\Psi\right\rangle =\zeta q\left|\boldsymbol\Psi\right\rangle \,,\;\;\zeta=\pm1.\label{4.1}\end{equation}
 We note that the spectrum of the QFT Hamiltonian $\hat{H}^{\mathrm{QFT}}$,\begin{equation}
\hat{H}^{\mathrm{QFT}}=\sum_{n}\varepsilon_{+,n}a_{n}^{+}a_{n}+\sum_{\alpha}\varepsilon_{+,\alpha}^{c}b_{\alpha}^{+}b_{\alpha}\,,\label{a19}\end{equation}
 in the one-particle sector reproduces that of particles and antiparticles
without an infinite number of negative-energy levels. A state vector
of QFT in a given moment of time $x^{0}$ will be denoted as $\left|\boldsymbol\Psi(x^{0})\right\rangle $.
This vector evolves in time according to the Schrödinger equation
\[
i\partial_{0}\left|\boldsymbol\Psi\left(x_{0}\right)\right\rangle =\hat{H}^{\mathrm{QFT}}\left|\boldsymbol\Psi\left(x_{0}\right)\right\rangle ,\]
 and remains in the one-particle sector due the fact that the Hamiltonian
$\hat{H}^{\mathrm{QFT}}$ commutes with the particle number operator\begin{equation}
\hat{N}=\sum_{n}a_{n}^{+}a_{n}+\sum_{\alpha}b_{\alpha}^{+}b_{\alpha}\,.\label{a21}\end{equation}

Let us introduce a coordinate representation of the Fock space of
time-dependent one-particle states. To this end consider the decompositions
of the field operators $\hat{\Psi}(x)$ and $\hat{\Psi}^{c}(x)$ as
\[
\hat{\Psi}\left(x\right)=\hat{\Psi}_{\left(-\right)}\left(x\right)+\hat{\Psi}_{\left(+\right)}\left(x\right)\,,\;\;\;\hat{\Psi}^{c}\left(x\right)=\hat{\Psi}_{\left(-\right)}^{c}\left(x\right)+\hat{\Psi}_{\left(+\right)}^{c}\left(x\right)\,,\]
 where $\hat{\Psi}^{c}$ is the charge-conjugated Heisenberg operator
of the field $\hat{\Psi}$, defined by $\hat{\Psi}_{\left(\pm\right)}^{c}=\left(\hat{\Psi}_{\left(\mp\right)}^{+}\Gamma^{2}\right)^{\mathrm{T}}$,
while the plus and minus terms are given by \begin{align*}
\hat{\Psi}_{\left(-\right)}\left(x\right) & =\sum_{n}a_{n}\psi_{+,n}\left(x\right)\,,\,\,\hat{\Psi}_{\left(+\right)}=\sum_{\alpha}b_{\alpha}^{+}\psi_{-,\alpha}\left(x\right)\,,\\
\hat{\Psi}_{\left(-\right)}^{c}\left(x\right) & =\sum_{\alpha}b_{\alpha}\psi_{+,\alpha}^{c}\left(x\right)\,,\,\,\hat{\Psi}_{\left(+\right)}^{c}\left(x\right)=\sum_{n}a_{n}^{+}\psi_{-,n}^{c}\,\left(x\right).\end{align*}
 With the help of these operators, we introduce the wave functions\[
\psi_{+}\left(x\right)=\left\langle 0\right|\hat{\Psi}_{\left(-\right)}\left(x\right)\left|\boldsymbol\Psi\left(0\right)\right\rangle ,\;\;\;\psi_{+}^{c}\left(x\right)=\left\langle 0\right|\hat{\Psi}_{\left(-\right)}^{c}\left(x\right)\left|\boldsymbol\Psi\left(0\right)\right\rangle .\]
 One can see that such wave functions specify completely the state
$\left|\boldsymbol\Psi\left(x_{0}\right)\right\rangle $. Since $\left|\boldsymbol\Psi\left(x_{0}\right)\right\rangle $
belongs either to the particle subspace $\mathcal{R}_{10}^{\mathrm{QFT}}$
or to the antiparticle subspace $\mathcal{R}_{01}^{\mathrm{QFT}}$,
we can define its four-component coordinate representation $\boldsymbol\psi\left(x\right)$
for each inequivalent representation of the gamma-matrices as \begin{equation}
\boldsymbol\psi_{+}\left(x\right)=\left(\begin{array}{c}
\psi_{+}\left(x\right)\\
0\end{array}\right),\;\;\;\boldsymbol\psi_{+}^{c}\left(x\right)=\left(\begin{array}{c}
0\\
\psi_{+}^{c}\left(x\right)\end{array}\right),\label{4.2}\end{equation}
 where $\psi\left(x\right)=\psi^{(\theta)}\left(x\right)$, and $\psi^{c}\left(x\right)=\psi^{(\theta)c}\left(x\right)$
in the $\Gamma_{\theta}$ representation. Using the projection operator
to the one-particle sector, \[
\int\left(\hat{\Psi}^{\dagger}\left|0\right\rangle \left\langle 0\right|\hat{\Psi}+\hat{\Psi}^{c\dagger}\left|0\right\rangle \left\langle 0\right|\hat{\Psi}^{c}\right)d\mathbf{x}=I_{2},\]
 we can present the QFT inner product $\left\langle \boldsymbol\Psi|\boldsymbol\Psi^{\prime}\right\rangle $
in terms of representatives, \[
\left\langle \boldsymbol\Psi|\boldsymbol\Psi^{\prime}\right\rangle =\left\{ \begin{array}{c}
\left(\psi_{+},\psi_{+}^{\prime}\right)\,,\;\zeta=+1\\
\left(\psi_{+}^{c},\psi_{+}^{c\prime}\right)\,,\;\zeta=-1\end{array}\right..\]
 It is easy to see that the equations \[
\hat{H}^{\mathrm{QFT}}\left|\Psi_{n}\right\rangle =\varepsilon_{+,n}\left|\Psi_{n}\right\rangle ,\;\;\;\hat{H}^{\mathrm{QFT}}\left|\Psi_{\alpha}^{c}\right\rangle =\varepsilon_{+,\alpha}^{c}\left|\Psi_{\alpha}^{c}\right\rangle ,\]
 with $\left|\Psi_{n}\right\rangle =a_{n}^{+}\left|0\right\rangle $
and $\left|\Psi_{\alpha}^{c}\right\rangle =b_{\alpha}^{+}\left|0\right\rangle $,
are written in the coordinate representation as follows \[
\hat{H}\boldsymbol\psi_{+,n}=\varepsilon_{+,n}\boldsymbol\psi_{+,n}\,,\;\;\;\hat{H}\boldsymbol\psi_{+,\alpha}^{c}=\varepsilon_{+,\alpha}^{c}\boldsymbol\psi_{+,\alpha}^{c}\,,\]
 so that for each inequivalent representation of the gamma-matrices
the Hamiltonian $\hat{H}$ can be identified with \begin{equation}
\hat{H}=\mathrm{bdiag}\left(\hat{h},\hat{h}^{c}\right),\label{4.3}\end{equation}
 and \begin{equation}
\boldsymbol\psi_{+,n}=\left(\begin{array}{c}
\psi_{+,n}\\
0\end{array}\right),\;\;\;\boldsymbol\psi_{+,\alpha}^{c}=\left(\begin{array}{c}
0\\
\psi_{+,\alpha}^{c}\end{array}\right).\label{4.4}\end{equation}

It is clear that in the coordinate representation the charge operator
$\hat{Q}^{\mathrm{QFT}}$ acts as the charge operator in the space
$\mathcal{R}_{\mathrm{ph}}^{\mathrm{eff}}$ of the QM, \begin{equation}
q\hat{\zeta}\boldsymbol\psi_{+}\left(x\right)=q\boldsymbol\psi_{+}\left(x\right),\;\;\; q\hat{\zeta}\boldsymbol\psi_{+}^{c}\left(x\right)=-q\boldsymbol\psi_{+}^{c}\left(x\right),\label{4.5}\end{equation}
 where $\hat{\zeta}$ is given by (\ref{3.36}).

Thus, the states (\ref{4.4}) form a basis in the coordinate representation
of $\mathcal{R}_{\mathrm{ph}}^{\mathrm{QFT}}$. These states are eigenstates
of the QFT charge operator, and \begin{equation}
\left(\boldsymbol\psi_{+,n},\boldsymbol\psi_{+,m}\right)=\left(\psi_{+,n},\psi_{+,m}\right)\,,\;\;\left(\boldsymbol\psi_{+,\alpha}^{c},\boldsymbol\psi_{+,\beta}^{c}\right)=\left(\psi_{+,\alpha}^{c},\psi_{+,\beta}^{c}\right)\,,\;\left(\boldsymbol\psi_{+,n},\boldsymbol\psi_{+,\alpha}^{c}\right)=0.\label{4.6}\end{equation}

We see that the Hamiltonian (\ref{4.3}) in the coordinate representation
of the one-particle sector of the $2+1$ QFT is precisely the Hamiltonian
(\ref{3.35}) for a fixed representation $\theta=+1$ (or $\theta=-1$)
of the gamma-matrices. Moreover, the one-particle sector of the physical
state space $\mathcal{R}_{\mathrm{ph}}^{\mathrm{QFT}}$ coincides
with the effective physical sector $\mathcal{R}_{\mathrm{ph}}^{\mathrm{eff}}$
defined by (\ref{3.34}). Thus, in backgrounds which do not violate
vacuum stability, the QM with an appropriate definition of the Hilbert
space can be identified with the one-particle sector of QFT. 

To describe the one-particle sector with both species of particles
(i.e., corresponding to $\theta=+1$ and $\theta=-1$), one needs
to use the four-component wave functions. Such wave functions (representatives
of $\left|\boldsymbol\Psi^{(\theta)}\left(x_{0}\right)\right\rangle $)
have the form \begin{align*}
\boldsymbol\Phi_{(\theta)+}\left(x\right) & =\left(\begin{array}{c}
\boldsymbol\psi_{(\theta)+}\left(x\right)\\
0\end{array}\right),\;\;\;\boldsymbol\Phi_{(\theta)+}^{c}\left(x\right)=\left(\begin{array}{c}
0\\
\boldsymbol\psi_{(\theta)+}^{c}\left(x\right)\end{array}\right),\\
\boldsymbol\psi_{(+1)+}\left(x\right) & =\left(\begin{array}{c}
\psi_{+}^{(+1)}\left(x\right)\\
0\end{array}\right),\;\;\;\boldsymbol\psi_{(-1)+}\left(x\right)=\left(\begin{array}{c}
0\\
\sigma^{1}\psi_{+}^{(-1)}\left(x\right)\end{array}\right),\\
\boldsymbol\psi_{(+1)+}^{c}\left(x\right) & =\left(\begin{array}{c}
\psi_{+}^{(+1)c}\left(x\right)\\
0\end{array}\right),\;\;\;\boldsymbol\psi_{(-1)+}^{c}\left(x\right)=\left(\begin{array}{c}
0\\
\sigma^{1}\psi_{+}^{(-1)c}\left(x\right)\end{array}\right),\end{align*}
 where $\psi_{+}^{(\theta)}\left(x\right)$ and $\psi_{+}^{(\theta)c}\left(x\right)$
are two-component representatives defined by equation (\ref{4.2})
for the $\theta$-representation of the gamma-matrices. In this coordinate
representation, the charge operator $\hat{Q}^{\mathrm{QFT}}$ acts
as follows \begin{equation}
q\hat{\zeta}\boldsymbol\Phi_{(\theta)+}\left(x\right)=q\boldsymbol\Phi_{(\theta)+}\left(x\right),\;\; q\hat{\zeta}\boldsymbol\Phi_{(\theta)+}^{c}\left(x\right)=-q\boldsymbol\Phi_{(\theta)+}^{c}\left(x\right)\,,\label{S8}\end{equation}
 where $\hat{\zeta}=\mathrm{bdiag}\left(I_{4},-I_{4}\right)$.

In this one-particle sector we can introduce the following spin polarization
operator \begin{equation}
\hat{S}^{\mathrm{QFT}}=\frac{1}{2}\int:\hat{\boldsymbol{\psi}}^{+}\gamma^{0}\Sigma^{3}\hat{\boldsymbol{\psi}}:d^{2}x,\label{S4}\end{equation}
 where $\hat{\boldsymbol{\psi}}$ and $\hat{\boldsymbol{\psi}}^{+}$
are the linear superpositions of the quantized four-component Dirac
fields $\hat{\boldsymbol{\psi}}_{(\theta)}$ and $\hat{\boldsymbol{\psi}}_{(\theta)}^{+}$,
respectively%
\footnote{In the usual description of spin polarization in $2+1$ dimensions
one refers to the angular mometum in the rest frame (see, for example,
\cite{JacN91,GitSh}). The spin operator found in this way cannot
be conserved.%
}. The operator (\ref{S4}) is a scalar under the $2+1$ Lorentz transformations
and conserved in any external field. In the representation (\ref{3.14})
of the gamma-matrices, we have $\gamma^{0}\Sigma^{3}=\left(\begin{array}{cc}
I_{2} & 0\\
0 & -I_{2}\end{array}\right)$. Consequently, in order to be eigenvectors of this matrix, the fields$\hat{\boldsymbol{\psi}}_{(\theta)}$
can be selected as follows, \begin{equation}
\hat{\boldsymbol{\psi}}_{(+1)}=\left(\begin{array}{c}
\hat{\Psi}^{(+1)}\\
0\end{array}\right),\;\;\;\hat{\boldsymbol{\psi}}_{(-1)}=\left(\begin{array}{c}
0\\
\sigma^{1}\hat{\Psi}^{(-1)}\end{array}\right),\label{S5}\end{equation}
 where $\hat{\Psi}^{(\theta)}$ is the two-component field operator
defined in the $\theta$-representation of the gamma-matrices (\ref{a2}).
Thus, we obtain \begin{equation}
\hat{S}^{\mathrm{QFT}}=\frac{1}{2q}\left(\hat{Q}_{+1}^{\mathrm{QFT}}-\hat{Q}_{-1}^{\mathrm{QFT}}\right)\mathbf{,}\label{S6}\end{equation}
 where $\hat{Q}_{\theta}^{\mathrm{QFT}}$ is the $2+1$ charge operator
(\ref{a20}) in the corresponding representation of the gamma-matrices.
We can see that this construction is trivial only if either one of
the particle species $\theta=+1$ or $\theta=-1$ is present. If this
is the case, then the conserved $2+1$ QFT spin polarization operator
is simply proportional to another conserved QFT scalar, the charge.

One-particle states $\left|\boldsymbol\Psi\right\rangle =\left|\boldsymbol\Psi^{(\theta)}\right\rangle $
are eigenstates of the spin polarization operator (\ref{S6}), \begin{equation}
\hat{S}^{\mathrm{QFT}}\left|\boldsymbol\Psi^{(\theta)}\right\rangle =\theta\frac{\zeta}{2}\left|\boldsymbol\Psi^{(\theta)}\right\rangle ,\;\;\;\zeta=\pm1.\label{S7}\end{equation}
 The species of a one-particle state is uniquely determined by the
operator $\hat{S}^{\mathrm{QFT}}$ times the charge of the state.

We can see that this operator is non-trivial in the $2+1$ extended
(four-component spinor) representation (\ref{S4}), since $\gamma^{0}\Sigma^{3}$
is not a unit matrix. The relation between the $2+1$ extended representation
and the two-component spinor representation of the Dirac QFT is similar
to the relation between the physical subspace $\mathcal{R}_{\mathrm{ph}}$
and the effective physical subspace $\mathcal{R}_{\mathrm{ph}}^{\mathrm{eff}}$,
discussed in Section 3. One can say that the space of two-component
spinors is an effective space of the $2+1$ Dirac theory.

In the coordinate representation, the equation (\ref{S7}) has the
form \begin{equation}
\hat{s}\boldsymbol\Phi_{(\theta)+}\left(x\right)=\frac{\theta}{2}\boldsymbol\Phi_{(\theta)+}\left(x\right),\;\;\;\hat{s}\boldsymbol\Phi_{(\theta)+}^{c}\left(x\right)=-\frac{\theta}{2}\boldsymbol\Phi_{(\theta)+}^{c}\left(x\right),\label{S9}\end{equation}
 where $\hat{s}=\frac{1}{2}\mathrm{bdiag}\left(\gamma^{0}\Sigma^{3},-\gamma^{0}\Sigma^{3}\right)$.

In Subsection 3.4, we have defined the effective physical sector $\mathcal{R}_{\mathrm{ph}}^{\mathrm{eff}}$
of the QM as the space of states (\ref{3.34}). Equivalently, we can
realize the effective physical sector of the QM in the extended (eight-component)
representation $\mathcal{\tilde{R}}_{\mathrm{ph}}^{\mathrm{eff}}$
as the space of states $\boldsymbol\Phi_{(\theta)+}$ and $\boldsymbol\Phi_{(\theta)+}^{c}$.
It is easy to see that the representation of the operator $\hat{s}$
in (\ref{S9}) is a representative of the operator $\hat{S}$ in $\mathcal{R}_{\mathrm{ph}}$,
namely, \[
\left(\boldsymbol\Psi_{\theta,+}\,,\hat{S}\boldsymbol\Psi_{\theta,+}^{\prime}\right)=\left(\boldsymbol\Phi_{(\theta)+},\hat{s}\boldsymbol\Phi_{(\theta)+}^{\prime}\right),\;\;\left(\boldsymbol\Psi_{\theta,+}^{c}\,,\hat{S}\boldsymbol\Psi_{\theta,+}^{\prime c}\right)=\left(\boldsymbol\Phi_{(\theta)+}^{c},\hat{s}\boldsymbol\Phi_{(\theta)+}^{\prime c}\right).\]
 Thus, we conclude that the physical subspace $\mathcal{\tilde{R}}_{\mathrm{ph}}^{eff}$
of the QM is identical with the one-particle sector of the $2+1$
spinor QFT in the four-component spinor representation.

\section{Discussion}

In this paper, we have quantized a $P$- and $T$-noninvariant pseudoclassical
model of a massive relativistic spin-$1/2$ particle in $2+1$ dimensions,
on the background of an arbitrary $U(1)$ gauge vector field. A peculiar
feature of the model at the classical level is that it contains a
bifermionic first-class constraint, which does not admit gauge-fixing.
It is shown that this first-class constraint can be realized at the
quantum level as a compact spectrum operator, which is imposed as
a condition on the state vectors (by analogy with the Dirac quantization
method). This allows us to generalize the quantization scheme \cite{GavGi00a,GavGi00b,GavGi01}
in case there is a bifermionic first-class constraint.

In doing so, we encounter the phenomenon of quantizing classical constants,
characteristic of pseudoclassical models. One ought to say that there
are different viewpoints concerning the nature of classical constants
in pseudoclassical models and their quantization \cite{GitGoTy,GitTy97,CPV93,bifermionic,Hov,Konst}.
One of these viewpoints \cite{CPV93,Konst} consists of replacing
the classical constants by dynamical variables, prior to quantization,
thus modifying the action. We think this approach is unnecessary,
and refer to \cite{GitTy02} for a thorough discussion. One may point
out that the restraint on the range of values of a certain parameter
is commonplace in quantum theory, and it is related to the details
of the quantization and operator ordering. There is no reason one
cannot treat classical constants in a similar manner, and restrict
the range of their values at the classical level. In addition, one
can also expect that the nature of these constants should change in
the passage to quantum theory, which is the case for dynamical variables.
In our model, on the classical level, $\theta$ is a bifermionic constant.
In course of quantization it passes into a constant matrix, whose
possible eigenvalues are defined by the quantum dynamics. In any case,
the role of classical constants, in particular the parameter $\theta$,
may be better clarified in the context of the semiclassical limit
of the QM, the discussion of which is presented in the Appendix. Another
viewpoint regarding the status of classical constants stems from the
understanding that the sole purpose of pseudoclassical models is to
provide a quantum theory. Therefore, one could fix the values of the
parameters right at the beginning to be precisely those values which
the QM dictates. In this sense, in the present model, one could interpret
the parameter $\theta$ as a real number already in the pseudoclassical
action.

We present a detailed construction of the Hilbert space and verify
that the constructed QM possesses the necessary symmetry properties.
We show that the condition of preservation of the classical symmetries
under the restricted Lorentz transformations and the $U(1)$ transformations
allows one to realize the operator algebra in an unambiguous way.
Within the constructed relativistic QM, we select a physical subspace
which describes the one-particle sector. The obtained realization
of the operator algebra differs significantly from the one obtained
in the quantization of a $P$-invariant pseudoclassical model of a
massive relativistic spin-$1/2$ particle in $2+1$ dimensions. The
physical sector of the QM contains both particles and antiparticles
with positive-energy $\hat{\Omega}$ levels, and exactly reproduces
the one-particle sector of the quantum theory of the $2+1$ spinor
field.

\subparagraph{{\large Acknowledgement}}

R.F., S.P.G. and P.Y.M are grateful to FAPESP. D.M.G. acknowledges
the support of FAPESP, CNPq and DAAD.

\section{Appendix. Semiclassical limit}

Let us prove that the quantum Hamiltonian $\underline{\hat{H}}$ (\ref{3.5})
provides a consistent realization of its classical analogue $\mathcal{H}_{\mathrm{eff}}$.
To this end, it is sufficient to show that the operator $\underline{\hat{\Omega}}$
has a correct semiclassical limit. For simplicity, we assume that
only a magnetic field is present. The contribution of an electric
field to the semiclassical limit of the operator $\underline{\hat{\Omega}}$
can be analyzed in complete analogy to the $3+1$ case, studied in
\cite{GavGi00b} (see the discussion of the contributions $\hat{\rho}_{1}$
and $\hat{\rho}_{2}$).

From the standard viewpoint accepted in quantum mechanics, the semiclassical
behavior of a wave packet corresponding to a particle takes place
when the packet is sufficiently well-localized in the phase space
of coordinates and momenta. Accordingly, it can be characterized by
the coordinates of the average position $\bar{x}^{k}$ and the average
momenta $\bar{p}_{k}$. At the same time, the mean square deviations
$\Delta x^{k}$ should be small in comparison with the characteristic
scale $L$ of the system in question, $\Delta x^{k}\ll L$, while
the mean square deviations $\Delta p_{k}$ should be small as compared
to $|p_{k}|$, $\Delta p_{k}\ll$ $|p_{k}|$. In accordance with these
conditions, the external field should be sufficiently homogeneous,
and change with time sufficiently slowly, so that it should not vary
considerably within distances commensurate with the size of a semiclassical
wave packet (SWP), while the SWP should not disperse within the time
interval of the observation\emph{.} Thus, it is sufficient to restrict
the analysis to the case of a time-independent field.

To prove that $\underline{\hat{\Omega}}=\mathrm{bdiag}\left(\hat{\Omega}_{+1},\hat{\Omega}_{-1}\right)$
is a consistent quantum realization of $\omega$ ($\omega=\omega_{0}$
in the absence of an electric field), it suffices to show that the
squared operator $\underline{\hat{\Omega}}^{2}=\left.\mathrm{bdiag}\left(\hat{\omega}_{0}^{2},\hat{\omega}_{0}^{2}\right)\right|_{x^{0}=\zeta\tau}$,
given by (\ref{3.11}), has a correct semiclassical limit, i.e., it
has the same expectation value on semiclassical states $\underline{\boldsymbol\Psi}_{\theta}^{cl}$
as the operator $\widehat{\underline{\Omega}^{2}}$, obtained by direct
quantization of $\omega_{0}^{2}$, \begin{equation}
\widehat{\underline{\Omega}^{2}}=\mathrm{bdiag}\left(\widehat{\omega_{0}^{2}},\widehat{\omega_{0}^{2}}\right),\;\;\;\widehat{\omega_{0}^{2}}=m^{2}+\left(\hat{p}_{k}+qA_{k}\right)^{2}-2iqF_{kl}\hat{\xi}^{k}\hat{\xi}^{l}.\label{ql1}\end{equation}
 Thus, we need to show that \begin{equation}
\left(\underline{\boldsymbol\Psi}_{\theta}^{cl}|\underline{\hat{\Omega}}^{2}\underline{\boldsymbol\Psi}_{\theta}^{cl}\right)=\left(\underline{\boldsymbol\Psi}_{\theta}^{cl}|\widehat{\underline{\Omega}^{2}}\underline{\boldsymbol\Psi}_{\theta}^{cl}\right),\label{job}\end{equation}
 where the inner product is defined by (\ref{3.3}).

In what follows, it is convenient to use the $x^{0}$-representation.
Thus, we pass from the vectors $\underline{\boldsymbol\Psi}_{\theta}^{cl}$
to the vectors $\boldsymbol\Psi_{\theta}^{cl}.$ The SWP is a superposition
of state vectors from the physical subspace $\mathcal{R}_{\mathrm{ph}}$,
i.e., we have either a particle SWP, or an antiparticle SWP. Let us
consider the case of a particle SWP (for charge-conjugated particles
the consideration is analogous) \[
\boldsymbol\Psi_{\theta}^{cl}(x)=\sum_{n}c_{+,n}^{(\theta)}\boldsymbol\Psi_{\theta,+,n}(x),\]
 where $c_{+,n}^{(\theta)}$ are the corresponding constant coefficients
of the decomposition of the SWP in eigenvectors. The coefficients
are defined at the initial time instant $x_{in}^{0}=0$ by the given
mean values of the coordinates $\mathbf{\bar{x}}_{in}$ and momenta
$\mathbf{\bar{p}}_{in}$ characterizing the SWP. The component structure
of semiclassical states $\boldsymbol\Psi_{\theta}^{cl}$ is that of
the states $\boldsymbol\Psi_{\theta}$ from the physical subspace
$\mathcal{R}_{\mathrm{ph}}$ (\ref{3.29}), \begin{align*}
 & \boldsymbol\Psi_{\theta}^{cl}=\left(\begin{array}{c}
\Psi_{\theta}^{cl}\\
0\end{array}\right),\;\;\Psi_{\theta}^{cl}=\frac{1}{\sqrt{2}}\left(\begin{array}{c}
\boldsymbol\psi_{\theta}^{cl}\\
\gamma^{0}\boldsymbol\psi_{\theta}^{cl}\end{array}\right),\\
 & \boldsymbol\psi_{+1}^{cl}=\left(\begin{array}{c}
\psi^{\left(+1\right)cl}\\
0\end{array}\right),\;\;\boldsymbol\psi_{-1}^{cl}=\left(\begin{array}{c}
0\\
\sigma^{1}\psi^{\left(-1\right)cl}\end{array}\right),\end{align*}
 where \[
\psi^{(\theta)cl}(x)=\sum_{n}c_{+,n}^{(\theta)}\psi_{+,n}^{(\theta)}(x).\]

As shown in Section 3, we can calculate the mean values of the function
$\mathcal{\tilde{F}}\left(\hat{X}^{k},\hat{P}_{k},\hat{H}_{x^{0}}\right)$
by using its representative $\mathrm{bdiag}\left(\tilde{f}\left(x^{k},\hat{p}_{k},\hat{h}_{\theta}\right),\tilde{f}^{c}\left(x^{k},\hat{p}_{k},\hat{h}_{\theta}^{c}\right)\right)$
in the effective physical subspace $\mathcal{R}_{\mathrm{ph}}^{\mathrm{eff}}$.
Thus, for example, the mean values of $\mathbf{x}$ and $\mathbf{\hat{p}}$
at the time instant $x^{0}$ can be expressed as follows, \begin{align*}
\mathbf{\bar{x}} & =\left(\boldsymbol\Psi_{\theta}^{cl}(x^{0}),\mathbf{\hat{X}}\boldsymbol\Psi_{\theta}^{cl}(x^{0})\right)=\left(\psi^{(\theta)cl}(x^{0}),\mathbf{x}\psi^{(\theta)cl}(x^{0})\right),\\
\mathbf{\bar{p}} & =\left(\boldsymbol\Psi_{\theta}^{cl}(x^{0}),\mathbf{\hat{P}}\boldsymbol\Psi_{\theta}^{cl}(x^{0})\right)=\left(\psi^{(\theta)cl}(x^{0}),\mathbf{\hat{p}}\psi^{(\theta)cl}(x^{0})\right).\end{align*}
 These mean values depend on the parameter $x^{0}$, as well as on
the initial values of $\mathbf{\bar{x}}_{in}$ and $\mathbf{\bar{p}}_{in}$:
$\mathbf{\bar{x}=\bar{x}}(x^{0},\mathbf{\bar{x}}_{in},\mathbf{\bar{p}}_{in})$
and $\mathbf{\bar{p}=\bar{p}}(x^{0},\mathbf{\bar{x}}_{in},\mathbf{\bar{p}}_{in})$.
In accordance with the above definition, we obtain the semiclassical
behavior of a wave packet in case the equations of motion for the
mean values to the leading order in the expansion obey the classical
equations of motion \begin{align*}
 & \frac{d\bar{x}^{k}}{dx^{0}}=\left\{ \bar{x}^{k},\epsilon_{\theta}(\mathbf{\bar{x}},\mathbf{\bar{p}})\right\} \;,\\
 & \frac{d(\bar{p}_{k}+qA_{k}(\mathbf{\bar{x}}))}{dx^{0}}=\left\{ \bar{p}_{k}+qA_{k}(\mathbf{\bar{x}}),\epsilon_{\theta}(\mathbf{\bar{x}},\mathbf{\bar{p}})\right\} \,,\end{align*}
 where $\epsilon_{\theta}(\mathbf{\bar{x}},\mathbf{\bar{p}})$ is
the mean value of the Hamiltonian (in a stationary background, it
does not depend on the time $x^{0}$) \[
\epsilon_{\theta}(\mathbf{\bar{x}},\mathbf{\bar{p}})=\left(\boldsymbol\Psi_{\theta}^{cl}(x^{0}),\mathbf{\hat{H}}\boldsymbol\Psi_{\theta}^{cl}(x^{0})\right)=\left(\psi^{(\theta)cl}(x^{0}),\hat{h}^{(\theta)}\psi^{(\theta)cl}(x^{0})\right)\;.\]
 Then we can see that thus determined SWP satisfies (with semiclassical
accuracy) the evolution equation \begin{align*}
 & i\hbar\partial_{0}\psi^{(\theta)cl}(x)=\hat{\epsilon}_{\theta}\psi^{(\theta)cl}(x)\,,\\
 & \hat{\epsilon}_{\theta}=\epsilon_{\theta}(\mathbf{\bar{x}},\mathbf{\bar{p}})+\frac{\partial\epsilon_{\theta}(\mathbf{\bar{x}},\mathbf{\bar{p}})}{\partial\bar{p}_{k}}\left(\hat{p}_{k}-\bar{p}_{k}\right)+\frac{\partial\epsilon_{\theta}(\mathbf{\bar{x}},\mathbf{\bar{p}})}{\partial\bar{x}^{k}}\left(x^{k}-\bar{x}^{k}\right).\end{align*}
 This spinor (with semiclassical accuracy) can be represented in the
form \[
\psi^{(\theta)cl}(x)=\left[\Gamma_{\theta}^{0}\hat{\epsilon}_{\theta}+\Gamma_{\theta}^{j}\left(i\hbar\partial_{j}-qA_{j}(\mathbf{x})\right)+m\right]\varphi^{(\theta)cl}(x),\]
 where the function \[
\varphi^{(\theta)cl}(x)=\exp\left\{ -\frac{i}{\hbar}\hat{\epsilon}_{\theta}x^{0}\right\} \varphi^{(\theta)cl}(\mathbf{x}),\]
 obeys the squared Dirac equation (with semiclassical accuracy) \begin{equation}
\left[\left(\hat{\epsilon}_{\theta}\right)^{2}-D\right]\varphi^{(\theta)cl}(\mathbf{x})=0\,,\; D=m^{2}+\left(i\hbar\partial_{k}-qA_{k}(\mathbf{x})\right)^{2}-\theta\hbar\sigma^{3}qB(\mathbf{x})\,,\,\, F_{21}=B\,.\label{sqD}\end{equation}
 Therefore, the explicit form of the mean energy reads \[
\epsilon_{\theta}(\mathbf{\bar{x}},\mathbf{\bar{p}})=\sqrt{m^{2}+\left(\bar{p}_{j}+qA_{j}(\mathbf{\bar{x}})\right)^{2}-\theta\hbar<\sigma^{3}>qB(\mathbf{\bar{x}})}\,\,,\]
 where $\left\langle \sigma^{3}\right\rangle $ is the corresponding
mean value of the matrix $\sigma^{3}$, which will be specified below.
We can see that a state with a given spin polarization, and providing
the necessary semiclassical limit, is obtained by choosing the column
$\varphi^{(\theta)cl}(\mathbf{x})$ in the form \[
\varphi^{(\theta)cl}(\mathbf{x})=\left(\begin{array}{c}
f^{(\theta)cl}(\mathbf{x})\\
0\end{array}\right).\]
 Then we finally have \[
\epsilon_{\theta}(\mathbf{\bar{x}},\mathbf{\bar{p}})=\sqrt{m^{2}+\left(\bar{p}_{j}+qA_{j}(\mathbf{\bar{x}})\right)^{2}-\theta\hbar qB(\mathbf{\bar{x}})}.\]
 It is exactly the function $\varphi^{(\theta)cl}(x)$ which defines
(to the leading order in the approximation) the form of the SWP as
a function of coordinates and momenta. This function can be represented
as a well-localized wave packet of solutions of the squared Dirac
equation, \[
\varphi^{(\theta)cl}(x)=\sum_{n}c_{+,n}^{(\theta)}\varphi_{+,n}^{(\theta)}(x),\]
 with the same coefficients $c_{+,n}^{(\theta)}$ as those for $\psi^{(\theta)cl}(x)$.

Using the SWP of a particle in a magnetic field, we have \[
\left(\boldsymbol\Psi_{\theta}^{cl},\hat{\Omega}\boldsymbol\Psi_{\theta}^{cl}\right)=\epsilon_{\theta}(\mathbf{\bar{x}},\mathbf{\bar{p}}),\]
 and, correspondingly, \[
\left(\boldsymbol\Psi_{\theta}^{cl},\hat{\Omega}^{2}\boldsymbol\Psi_{\theta}^{cl}\right)=\left(\epsilon_{\theta}(\mathbf{\bar{x}},\mathbf{\bar{p}})\right)^{2}.\]
 On the other hand, using the SWP we obtain the same result for the
mean value of the operator $\widehat{\Omega^{2}}$ defined by (\ref{ql1}).
This completes the proof of the fact that $\hat{\Omega}$ has a correct
semiclassical limit%
\footnote{It is easy to see that in the general case, when we have an electric
field satisfying the quasiclassical conditions (of being weak and
sufficiently homogeneous within the SWP), we obtain a similar result,
\[
\left(\boldsymbol\Psi_{\theta}^{cl},\hat{\Omega}^{2}\boldsymbol\Psi_{\theta}^{cl}\right)=\left(\epsilon_{\theta}(\mathbf{\bar{x}},\mathbf{\bar{p}})-qA_{0}(\mathbf{\bar{x}})\right)^{2}=m^{2}+\left(\bar{p}_{j}+qA_{j}(\mathbf{\bar{x}})\right)^{2}-\theta qB(\mathbf{\bar{x}}).\]
 Thus, the proof is also valid in the general case.%
}.

To finish the semiclassical analysis, we find it useful to make a
remark concerning the physical meaning of a semiclassical spinning
particle. It should be noted that the term $\hbar\sigma^{3}qB(\mathbf{x})$,
referred to as a quantum correction, usually is not included into
the expression for the classical energy. The reason for doing so is
the fact that in the homogeneous field (for a SWP) we have the term
$\left(\bar{p}_{j}+qA_{j}(\mathbf{\bar{x}})\right)^{2}\sim2|qB|\hbar n$,
with Landau level number $n\gg1$. Therefore, the term $\hbar qB(\mathbf{x})$,
giving a contribution commensurate with the difference of the energy
levels in the first summand, is negligibly small in the expression
for the semiclassical energy $\epsilon_{\theta}(\mathbf{\bar{x}},\mathbf{\bar{p}})$.
However, in a non-homogeneous field, without changing the conditions
of semiclassics (i.e., within the conditions that permit one to characterize
motion by classical coordinates and momenta), one can see the influence
of spin magnetic moment on the classical trajectory. This can be observed
from the equation \[
\frac{d(\bar{p}_{k}+qA_{k}(\mathbf{\bar{x}}))}{dx^{0}}=-qF_{kl}(\mathbf{\bar{x}})\frac{d\bar{x}^{l}}{dx^{0}}+\theta q\frac{\partial B(\mathbf{\bar{x}})}{\partial\bar{x}^{k}}\frac{\hbar}{2\epsilon_{\theta}(\mathbf{\bar{x}},\mathbf{\bar{p}})}.\]
 Here, the term containing the field gradient $\partial B(\mathbf{\bar{x}})/\partial\bar{x}^{k}$
is much smaller than the preceding one. However, the first term causes
acceleration, which is always perpendicular to the velocity of the
particle. At the same time, the fluctuations of momenta or coordinates
do not change anything in this respect, since the acceleration is
simply a consequence of the structure of the minimal interaction (current-potential).
The acceleration caused by the second term, although small, is directed
alongside the field gradient, i.e., its direction is not related to
the motion of the particle, and is not affected by the fluctuations
of coordinates and momenta, because this is a consequence of the structure
of the interaction between the field and the magnetic momenta. In
principle, this permits one to separate the types of motion caused
by different interactions. In the given case, the presence of the
second term affects the particle in the same way as a spinless particle
is affected by a weak electric field directed alongside the gradient
of a magnetic field. The possibility of observing the effect of the
interaction with the spin magnetic moment on a classical trajectory
implies that one deals with a classical theory of a spinning particle
in the usual sense: one can describe motion using the concept of a
trajectory.

\end{document}